\documentclass{emulateapj}
\usepackage{apjfonts}

\begin{document}

\newcommand{\sigmaoned}{\sigma_{\hbox{1D}}}
\newcommand{\sigmaninety}{\sigma_{90}}
\newcommand{\SNR}{SNR}
\newcommand{\FIRST}{{\sl FIRST\/}}

\title{The Last of \FIRST: The Final Catalog and Source Identifications}

\shorttitle{The Last of \FIRST}
\shortauthors{Helfand, White \& Becker}
\slugcomment{To appear in the Astrophysical Journal}

\author{David J. Helfand\altaffilmark{1,2}, 
  Richard L. White\altaffilmark{3}
  and Robert H. Becker\altaffilmark{4,5}}

\altaffiltext{1}{Department of Astronomy, Columbia University, 
  Mail Code 5233, Pupin Hall, 538 West 120th Street, 
  New York City, NY 10027.}
\altaffiltext{2}{Quest University Canada, 3200 University Blvd., Squamish, BC V8B 0N8  Canada}
\altaffiltext{3}{Space Telescope Science Institute, 
  3700 San Martin Drive, Baltimore, MD 21218.}
\altaffiltext{4}{Department of Physics, University of California, 
  Davis, 1 Shields Avenue, Davis, CA 95616-8677.}
\altaffiltext{5}{IGPP/Lawrence Livermore National Laboratory}
\email{djh@astro.columbia.edu}

\begin{abstract}

The \FIRST\ survey, begun over twenty years ago, provides the
definitive high-resolution map of the radio sky. This VLA survey
reaches a detection sensitivity of 1~mJy at 20~cm over a final
footprint of 10,575 deg$^2$ that is largely coincident with the
Sloan Digital Sky Survey area.  Both the images and a catalog
containing 946,432 sources are available through the \FIRST\ web
site (\url{http://sundog.stsci.edu}).  We record here the authoritative
survey history, including hardware and software changes that affect
the catalog's reliability and completeness. In particular, we use
recent observations taken with the JVLA to test various aspects of
the survey data (astrometry, CLEAN bias, and the flux density scale).
We describe a new, sophisticated algorithm for flagging potential
sidelobes in this snapshot survey, and show that fewer than 10\%
of the catalogued objects are likely sidelobes, and that these are
heavily concentrated at low flux densities and in the vicinity of
bright sources, as expected. We also report a comparison of the
survey with the NRAO VLA Sky Survey (NVSS), as well as a match of
the \FIRST\ catalog to the SDSS and 2MASS sky surveys. The NVSS
match shows very good consistency in flux density scale and astrometry
between the two surveys. The matches with 2MASS and SDSS indicate
a systematic $\sim10-20$~mas astrometric error with respect to the
optical reference frame in all VLA data that has disappeared with
the advent of the JVLA. We demonstrate strikingly different behavior
between the radio matches to stellar objects and  to galaxies in
the optical and IR surveys reflecting the different radio populations
present over the flux density range 1--1000~mJy.
As the radio flux density declines, stellar counterparts (quasars)
get redder and fainter, while galaxies get brighter and have colors
that initially redden but then turn bluer near the \FIRST\ detection limit.

Implications for
future radio sky surveys are also briefly discussed. In particular, we
show that for radio source identification at faint optical magnitudes,
high angular resolution observations are essential, and cannot be
sacrificed in exchange for high signal-to-noise data. The value of
a JVLA survey as a complement to SKA precursor surveys is briefly
discussed.

\end{abstract}

\keywords{catalogs --- methods: data analysis, statistical --- radio
continuum: general --- surveys}

\section{Introduction}\label{intro}

Faint Images of the Radio Sky at Twenty-centimeters were collected
over a period of eighteen years at the (now, Jansky) Very Large
Array.  The original proposal to use the world's premiere radio
telescope for the relatively mundane task of surveying the sky was
submitted in August of 1990. Pilot observations for the \FIRST\
survey began in April of 1993, fifty years to the month after Grote
Reber's original radio sky survey \citep{reber1944}. The final observations
were completed in the Spring of 2011 in an expansion of the survey to cover the
SDSS3 sky survey area. Over this extended period, a
number of VLA hardware changes were implemented, not least being
the transformation of the entire array into the JVLA. In addition, a number
of changes to the data reduction software and processing algorithms
also took place over the course of the survey. Furthermore, the
lower-resolution NVSS survey was completed during this interval,
as were sky surveys at infrared (2MASS) and optical (SDSS) wavelengths.

In this paper, we describe the final products of the \FIRST\
survey.  The \FIRST\ images and catalogs are distributed through
the \FIRST\ web site\footnote{\url{http://sundog.stsci.edu}}, the Mikulski
Archive for Space Telescopes\footnote{\url{http://archive.stsci.edu/vlafirst/}} and other
archival sites.

A history of the hardware and software changes during the
project is provided to alert users to possible systematics in the
survey images and catalog (\S\ref{history}); we also note various rare anomalies
in the VLA system that have been discovered as a consequence of
examining over 75,000 snapshot images. In some of this work, we
make use of matches of the \FIRST\ catalog to other all-sky
surveys, the science from which is reported later in the paper.
We assess the astrometric accuracy of the \FIRST\ catalog,
which reveals some small systematic offsets in the positions
(\S\ref{astrometry}).  We
go on to describe the algorithm developed to assign to each source
a probability that it is a spurious catalog entry resulting from a
sidelobe of a bright source elsewhere in the field (\S\ref{sidelobes}). Various
tests of this algorithm using radio and optical catalog matching
provide an assessment of the algorithm's reliability.

\begin{figure*}
\includegraphics[width=\linewidth]{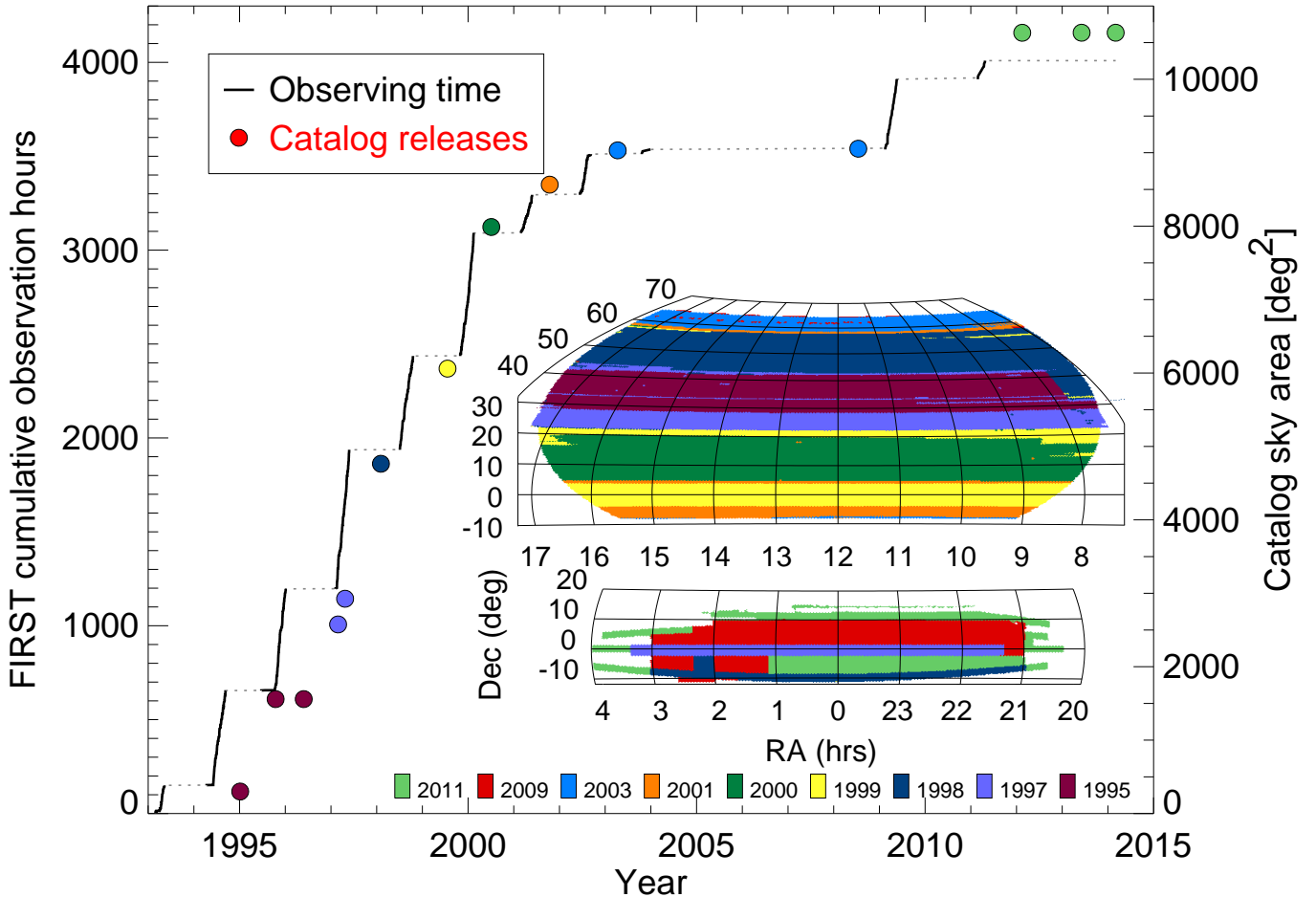}
\caption{The twenty-year history of the \FIRST\ survey showing
the sky coverage (nearly vertical dark lines) as a function of time.
The right axis shows the sky area in square degrees and the left
axis shows the cumulative observing time.  The insert shows the sky
coverage color-coded by the catalog release dates, which are also
indicated by circles colored to match the coverage map.  The final project,
representing an investment of 4009 hours of VLA time, generated a
survey covering 10,575 deg$^2$ of sky. \label{coverage}}
\end{figure*}

We compare the \FIRST\ survey catalog with the results of
three other major sky surveys that cover most or all of the same
$> 10,000$ deg$^2$ of sky: the NVSS \citep{condon1998}, the 2MASS
survey \citep{skrutskie2006}, and the SDSS \citep{york2000}
plus its more recent extensions \citep{ahn2014}.
In particular,
we use the NVSS match (\S\ref{NVSS}) to quantify the accuracy of
the \FIRST\ flux density scale and the degree to which faint
sources are missed as a consequence of the increased noise level
near bright sources, as well as the incompleteness of \FIRST\
for very extended objects. The 2MASS (\S\ref{2MASS}) and SDSS (\S\ref{SDSS}) matches
provide information on the astrometric accuracy of the respective
surveys, as well as insight into the classes of sources that populate
the radio sky at milliJansky flux densities. We conclude with a
commentary (\S\ref{conclusions}) on the survey's utility and on the lessons learned from this undertaking
that can be used to inform the next generation of radio sky surveys. In particular, we examine critically the effect of survey resolution on optical/IR identification programs for radio sources.

Readers uninterested in the detailed technical history and/or the
subtle catalog biases useful only for those using \FIRST\ in
large, statistical surveys, should skip to the science results, which
begin in \S\ref{NVSS}.

\begin{deluxetable}{lcccc}
\tablecolumns{5}
\tablewidth{0pt}
\tablecaption{Summary of \FIRST\ Catalog Releases\label{table-catalog-releases}}
\tablehead{
\colhead{Release Date} &  \colhead{Number of}   &  \colhead{North Galactic} & \colhead{South Galactic} & \colhead{Total} \\
\colhead{} &  \colhead{Sources}   &  \colhead{Cap Area} & \colhead{Cap Area} & \colhead{Area} \\
\colhead{} &  \colhead{}   &  \colhead{(deg$^2$)} & \colhead{(deg$^2$)} & \colhead{(deg$^2$)} \\
\colhead{(1)} & \colhead{(2)} & \colhead{(3)} & \colhead{(4)} & \colhead{(5)}
}
\startdata
1995 Jan 06 & \phn26,892 & \phn300 & \phn\phn0     & \phn300 \\
1995 Oct 16 &    138,665 &    1559 & \phn\phn0     &    1559 \\
1996 May 28 &    138,665 &    1559 & \phn\phn0     &    1559 \\
1997 Feb 27 &    236,040 &    2576 & \phn\phn0     &    2576 \\
1997 Apr 24 &    268,047 &    2576 &       350     &    2926 \\
1998 Feb 04 &    437,429 &    4153 &       611     &    4764 \\
1999 Jul 21 &    549,707 &    5448 &       611     &    6060 \\
2000 Jul 05 &    722,354 &    7377 &       611     &    7988 \\
2001 Oct 15 &    771,076 &    7954 &       611     &    8565 \\
2003 Apr 11 &    811,117 &    8422 &       611     &    9033 \\
2008 Jul 16 &    816,331 &    8444 &       611     &    9055 \\
2012 Feb 16 &    946,464 &    8444 &      2191\phn &   10635\phn \\
2013 Jun 05 &    971,268 &    8444 &      2191\phn &   10635\phn \\
2014 Mar 04\tablenotemark{a} & 946,432 &  8444 & 2131\phn & 10575\phn \\
2014 Dec 17\tablenotemark{a} & 946,432 &  8444 & 2131\phn & 10575\phn
\enddata
\tablenotetext{a}{These catalog versions have fewer sources and
a slightly reduced sky area because the JVLA images
are not co-added with older VLA images.  See text for more
details.}
\end{deluxetable}

\section{The Survey History}\label{history}

The VLA\footnote{The Karl G. Jansky Very Large Array is an instrument of the National Radio Astronomy
Observatory, a facility of the National Science Foundation operated
under cooperative agreement by Associated Universities, Inc.}
pilot observations in 1993 aided us in designing the pointing
grid for the survey (also adopted by the NVSS) and in developing
the basic data reduction algorithms for turning snapshot visibilities
into final survey images. These basic attributes of the project are
described in detail in \cite{becker1995} and the catalog,
constructed as detailed in \cite{white1997}; those papers should
remain the primary references when making use of \FIRST\ results.
Here, for the record, we document the technical history of the
project and describe all factors that might affect the character
and integrity of the data products.

In Figure~\ref{coverage}, we show the time allocated and the sky
coverage achieved over the 18 years of survey observations. A total
of eleven observing sessions in the B-configuration led to the
accumulation of just over 4000 hours (5.5 months) of observing time
that resulted in 10,575 deg$^2$ of sky coverage (8444 deg$^2$ in
the north Galactic cap and 2131 deg$^2$ in the southern cap). A
coverage map, color-coded by observation epoch, is displayed in the
inset to Figure 1; the catalog-release points are colored to match
the relevant coverage areas.

As originally envisioned, all data were released to the public
archive on the day they were taken, and all images were fully reduced
and put on a public website before a new set of observations
commenced. More than a dozen catalog releases were issued; they are located
in Figure 1 by date and sky coverage (right axis).
A summary of the catalog releases is given in
Table~\ref{table-catalog-releases}.
The final 2014~December~17 release contains 946,432 entries.
A history of catalog format changes and error corrections can be
found online\footnote{\url{http://sundog.stsci.edu/first/catalogs/history.html}}.

\subsection{Hardware and observing mode changes}\label{hardware}

The only major hardware change over the period of the survey was
occasioned by the transition of the array from its original
configuration to the JVLA. During the 2009 observing session, roughly
half the array had the original 20~cm receivers, while half had been
changed to the JVLA receivers. The center frequencies of the two
IFs were held constant, but the shapes of the receiver bandpasses
were different. In the 2011 session, the JVLA transition was complete.
We observed at a different center frequency (1335 MHz vs. two narrow
bands centered at 1365~MHz and 1435~MHz) with a 128~MHz total
bandwidth comprised of 64 2-MHz channels (vs. $2 \times 7$ 3-MHz
channels for the remainder of the survey); data taken in a second
128-MHz band centered at 1730~MHz, to be used in deriving source
spectral indices, will be reported elsewhere.  All the changes in
2011 were motivated by a combination of the changing 
requirements of the JVLA receiver/correlator system, plus the desire to
reduce the observing time and to extract additional science using
the new capabilities of the JVLA.

While the pilot observations used a 3-second integration time, we
quickly adopted 5-second integrations as standard. For the
2011 JVLA observations, however, the wide bandwidth 
required a reduction of the integration
time to 1 second.  These changes
affect the data analysis but do not lead to appreciable changes
in the survey data products.
The on-target dwell time was 165-seconds
per field until 2011, at which time the wider bandwidth available
allowed us to reach our 1~mJy sensitivity while reducing
the observing time to 1 minute per field.
The wider bandwidth and different frequency do have some effects
on the data (discussed further below).
The shorter integration time also required a change in the pointing
pattern, since the time spent per field was then much less than the
sidereal rate at which the sky passed overhead. A traveling-salesman
algorithm using simulated annealing \citep{kirkpatrick1983} was
developed to minimize slew time between fields.

During both 2009 and 2011, the Sun was located in the sky region
to be observed, requiring us to schedule around a zone of avoidance
centered on the apparent solar position. A 7-degree avoidance radius
was used in 2009; the zone was expanded to 10 degrees in 2011
owing to increased solar activity.  This, coupled with non-optimal
scheduling of our time allocations, led to many of the observations
being taken farther from the zenith than was desirable.  The scheduling
challenges combined with the loss of a significant quantity of data
to interference also resulted in some holes in the sky coverage and
led to the disconnected island of coverage that is visible in
Figure~\ref{coverage} near $\delta=15^\circ$ in the south Galactic
cap.

\begin{figure}
\includegraphics[width=\linewidth]{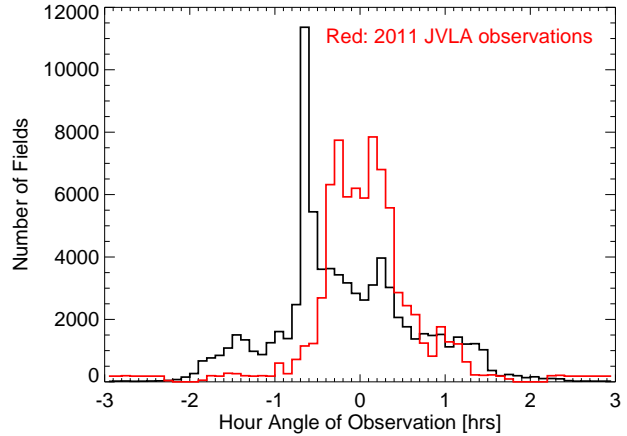}
\caption{Distribution of hour angles for the \FIRST\ observations.
The hour angle is zero for fields observed as they pass the meridian.
The black line shows the distribution for the entire survey.
The red line shows the distribution
for the 2011 JVLA observations, normalized to the same area as the overall
distribution. The latter distribution differs owing to shorter 1-minute integrations
and the need to avoid pointing toward the Sun.
\label{fig-hour-angles}}
\end{figure}

Figure~\ref{fig-hour-angles} shows how the observations were
distributed with respect to the meridian.  Ideally the fields
would all have been observed as they passed the meridian (zero
hour angle), but the real distribution is more complex.
90\% of the observations for the entire survey were acquired within 1.4 hours
of the meridian.  The distribution
is noticeably spiky and asymmetrical because the observing time blocks
allocated were often non-optimal, requiring that we
observe fields off the meridian.  The distribution for the 
short 1-minute JVLA observations, shown by the red line, is more symmetrical
(the result of our traveling salesman algorithm) but also has a very
extended tail at large hour angles (due to the need to avoid the
Sun.)  All these effects conspire to produce point-spread functions
(PSFs) and sidelobe patterns that vary slightly from year-to-year.
For most purposes, however, the PSF can be 
treated as if it were uniform.  Over the northern sky the PSF is a
circular Gaussian with FWHM 5.4\arcsec.  South of declination
${+}4^\circ33\arcmin21\arcsec$ 
the beam becomes elliptical, $6.4\arcsec\times5.4\arcsec$, with the major axis running
north-south.  South of ${-}2^\circ30\arcmin25\arcsec$ the elliptical beam
size increases further to $6.8\arcsec\times5.4\arcsec$.

In the final 2014Dec17 version of the catalog, the JVLA images
are not co-added with older VLA images to avoid problems resulting
from the different frequencies and noise properties of the
two datasets.  That leads to small gaps in sky coverage at the
boundaries between the JVLA and VLA regions, but has the advantage
that it cleanly separates sources from the old and new configurations.
All sources in the final catalog that have field names ending with
`W' come from the JVLA data.

\bigskip
\bigskip
\subsection{VLA anomalies}

Unsurprisingly, with 4000 hours of observing time to reduce
and a million sources to catalog, subtle and/or rare effects, not
noticeable in ordinary VLA programs, can appear. We summarize several
such anomalies here.

\textbf{Misassignment of data blocks:} Searches of the \FIRST\
database for source variability led to the discovery of a bug in
the online VLA software that appends to the current observation a
few integrations from the previous observation. This was revealed
when investigating the (ultimately spurious) detection of variability
in a bright source \citep{galyam2006,ofek2010}. All $uv$ datasets were
subsequently checked for this error and 190 cases of discontinuities
in $uvw$ values within a single scan were identified \citep{thyagarajan2011}.
These data sets were edited and the images re-made.
Note that all versions of the catalog prior to 2012 may contain a
few spurious sources as a consequence of this error.

\textbf{Misassignment of array configuration:} In collecting the
global observation records to include in this paper, we noticed that
the VLA archive sometimes records incorrect array configuration
labels for some observations. All observations for the \FIRST\
survey were, in fact, taken in the B configuration. The VLA archive
also is currently missing entries for $\sim1$\% of the \FIRST\ fields.

\textbf{Image stretch and rotation:} By comparing the positions of
sources observed in multiple pointings, we discovered that there
exist both a stretch and a rotation of VLA images, which we infer
are related to small clock errors at the VLA and small changes in
bandpass shape \citep[see footnote 6 of][]{white1997}. The astrometric
errors introduced are very small ($< 0.1\arcsec$) for most sources,
as the errors tend to cancel out in the co-added images. Only for
sources at the very edge of the survey coverage area can the errors
rise to $\sim 0.3\arcsec$.  For the first four epochs of observation,
we solved for the stretch and rotation corrections using multiply
observed sources. Since the parameters were quite stable from epoch
to epoch, we then fixed the correction and applied it to each epoch's
images through 2003. The last two epochs have not had these small
corrections applied. Thus, the positions of sources at the edge of
the coverage in the south Galactic cap should be assigned slightly
larger astrometric uncertainties.

\subsection{Data reduction changes}\label{data-reduction}

\begin{figure}
\includegraphics[width=\linewidth]{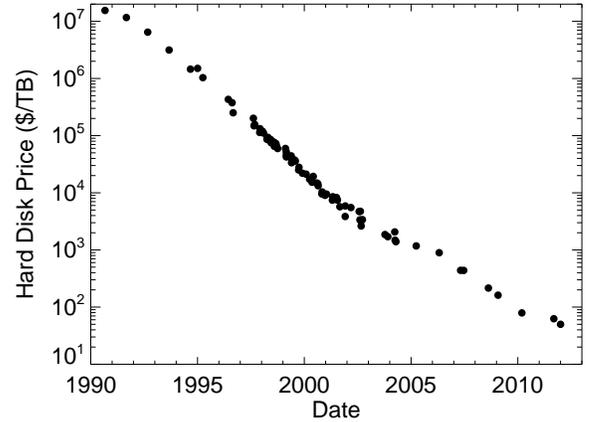}
\caption{Price of disk storage (dollars per terabyte, using constant
2012 dollars) over the duration of the \FIRST\ survey.  Storage prices
have dropped by a factor of roughly 300,000 since the \FIRST\ survey
proposal was submitted in 1990. Computer processing and memory have also
become dramatically less expensive over the lifetime of the survey.
\label{fig-disk-prices}}
\end{figure}

Computing power and storage capacity have increased dramatically
over the life of this project (Fig.~\ref{fig-disk-prices}).
In addition, the AIPS software package
has evolved, as have the scripts we run to process the \FIRST\
data. Finally, the recent upgrade to the JVLA hardware has required
changes in the processing pipeline. We briefly note here all
significant changes to the data processing.

\textbf{Data flagging:}
Prior to 2011, the data were collected in two sets of seven
3-MHz channels, and if interference was strong in one channel, all
seven channels of data for that integration period (5 seconds) were
deleted. With the advent of the broad-bandwidth JVLA correlator,
identified interference in one 2-MHz channel leads to the deletion
of that channel and the two adjacent channels only.

\textbf{Self-calibration:}
The self-calibration process \citep{hogbom1974} requires iteratively CLEANing images
and using the flux models from those images to improve the 
antenna phase (and sometimes amplitude) calibrations. During
the initial years of the \FIRST\ survey, the computing required
to construct a map was daunting.  To render the computing tractable,
we adopted an approach of
making an initial low-resolution map to find bright sources
in the image and then using only a set of small maps centered
on those
source positions for the self-calibration iteration \citep{becker1995}.  Following
the six-year hiatus in 2009, we simplified the process for new data by
repeatedly making full-field images during the self-calibration iteration as
well as for producing the final map.

\textbf{Wide-field bright-source mapping:}
Owing to processing speed limitations, our original
approach for the final map was to augment a $2048\times2048$ map covering
the central primary beam with small satellite
maps placed at the positions of nearby bright sources
(from single-dish catalogs)
that lie within $10\arcdeg$ of each field center
\citep[for details see][]{becker1995}.
Beginning in 2009, we simply
made $4096 \times 4096$ images and dispensed with special
treatment for far off-axis sources. With 1.8~arcsec pixels, these
images span 2.0 degrees, which is four times larger than the 20~cm FWHM primary
beam diameter.  Note that because these images are snapshot observations,
there is no need for any special treatment of 3-D sky effects in
the processing.  The 3-D distortions are removed in post-processing
by warping the map as described in \cite{becker1995}.

Also, in the first section of the survey, each
field center was shifted by up to 0.9\arcsec\ so that the
brightest source in the field fell at the precise center of a pixel.
We never found any evidence that this significantly
affected image quality and, since it complicated the image processing
and the final products (which otherwise had a predictable pixel grid),
we abandoned this procedure in processing the final two epochs.

\textbf{Source extraction:}
We used our AIPS source extraction program HAPPY \citep{white1997}
to identify and measure the properties of sources in the \FIRST\ images.
HAPPY evolved over the course of the project as a result of
minor enhancements and bug fixes.  To ensure a more uniform catalog,
in 2007 we reprocessed all the data using the current version of HAPPY.
That led to minor changes in the source lists.

\begin{figure}
\includegraphics[width=\linewidth]{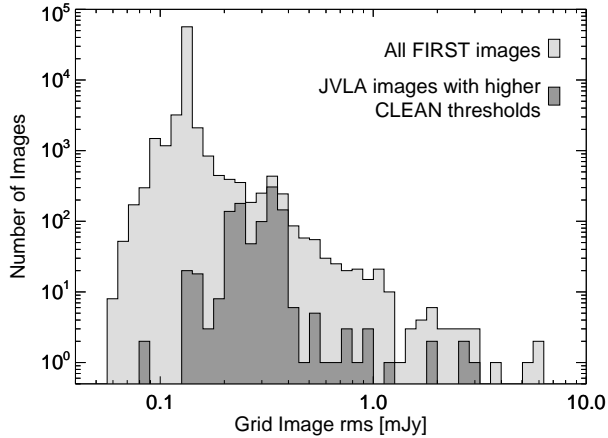}
\caption{Distribution of rms noise values in individual \FIRST\
survey images.  The dark shaded histogram
shows the noise distribution for JVLA fields observed on three days
with particularly bad interference; their mean noise levels are $\sim2.5$
times higher.
\label{fig-gridrms}}
\end{figure}

\begin{figure*}
\includegraphics[width=0.5\linewidth]{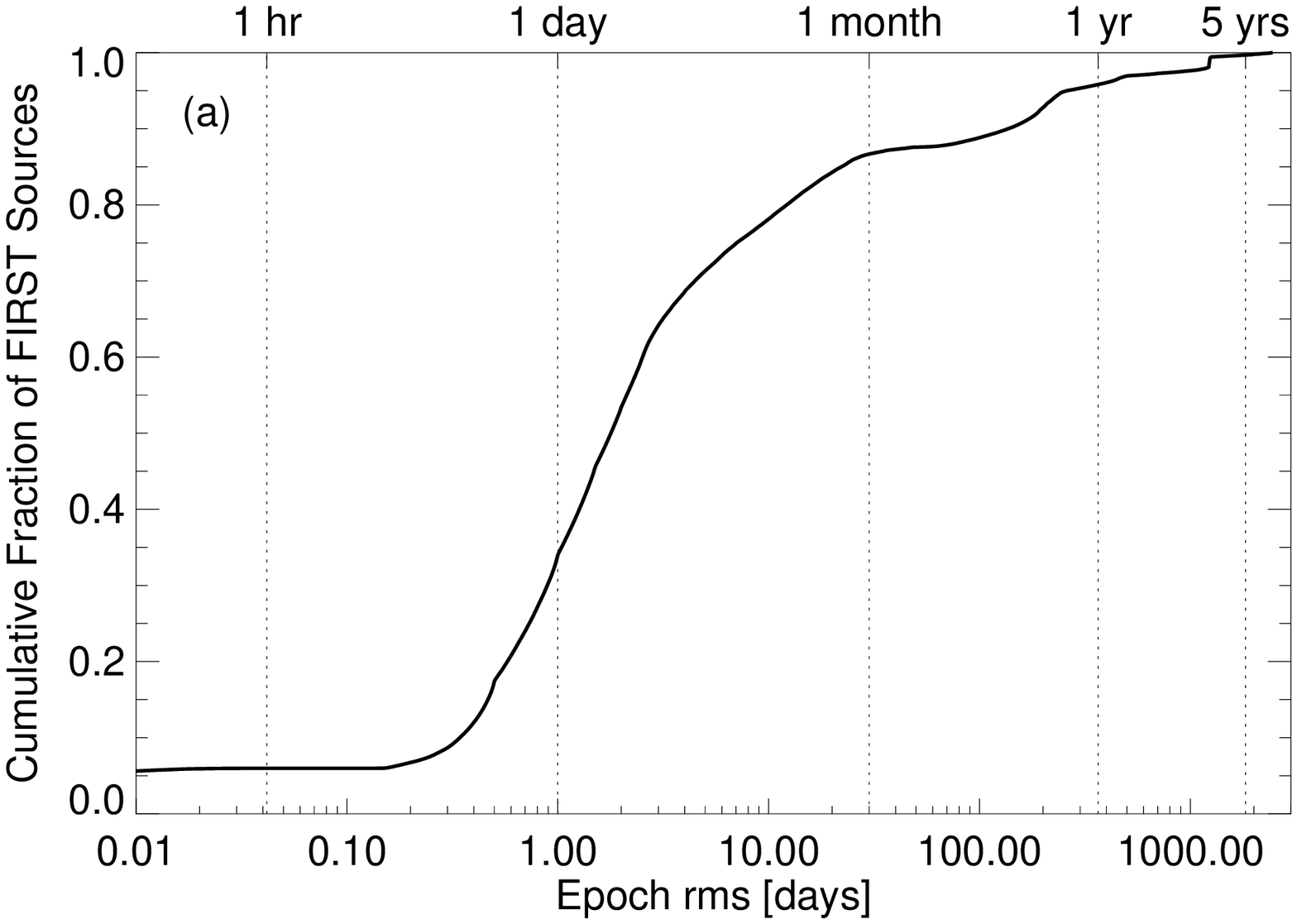}
\includegraphics[width=0.5\linewidth]{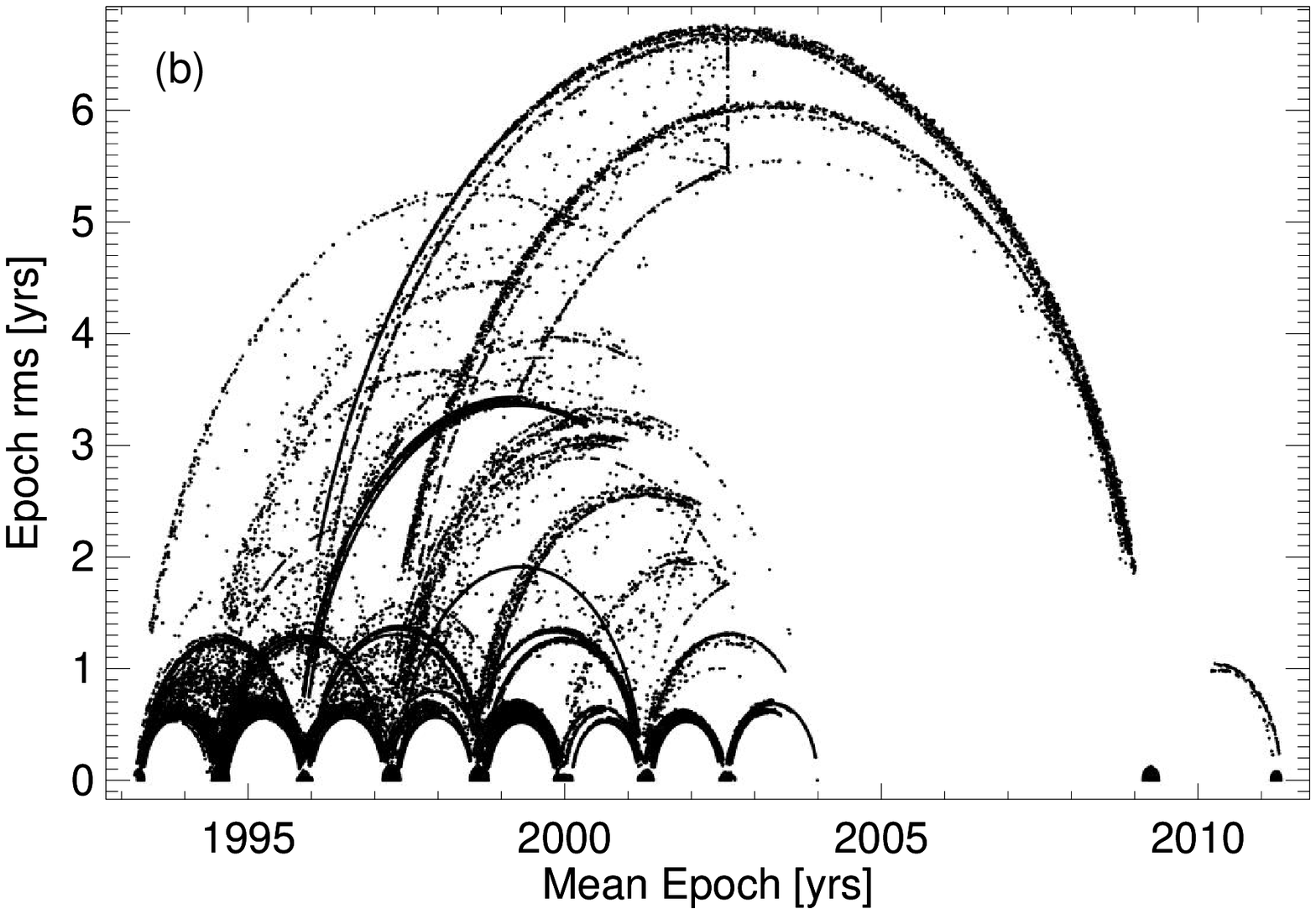}
\caption[Epoch distribution]{(a) Cumulative distribution of observation epoch
ranges $\sigma(t)$ in the \FIRST\ catalog. Many sources have flux densities measured over
an interval of an hour or less, but some come from combined observations that span years.
(b) The distribution of epoch rms with the mean epochs of the individual pointings.
The small arcs from year-to-year originate in the overlapping boundaries between
sky areas covered in consecutive observing seasons, as individual pointings from
different years are combined in the weighted, co-added images to produce
a rapidly changing mean epoch with a large rms value.
Larger epoch arcs and rms ranges come from overlaps between observing seasons that are
several years to decades apart; the longest arcs connect to the 2009 observing season,
which resumed observations of the south Galactic cap region after a long hiatus.
\label{fig-rmsdist}}
\end{figure*}

\textbf{Flux calibration and CLEAN bias:}
The flux density calibrators 3C286 and 3C48 were used for the North
and South Galactic Caps, respectively. The co-adding procedure to
create the final images from the individual grid images, as well
as the source detection and parameterization procedures, remained
unchanged throughout the survey period.

One of the systematic errors in VLA images that came to light from
\FIRST\ and NVSS was the so-called `CLEAN bias'. Sources extracted
from CLEANed snapshot images invariably lost flux in the CLEANing
process. Further analysis of the \FIRST\ images for studies of
sub-threshold sources through stacking \citep{white2007} showed
that the CLEAN bias effect is really a snapshot bias that reduces
the flux densities of even faint undetected sources.  This was
quantified by \cite{becker1995} and \cite{white2007} by
adding artificial sources into the UV data and comparing the input
and output flux densities for tens of thousands of point sources.
We continue to add a CLEAN bias offset of 0.25~mJy to the peak flux
densities of all sources \citep[see][]{becker1995}.

Since with the change of VLA hardware there was some concern that the
CLEAN bias would be different, we repeated these CLEAN bias tests
using artificial sources for each day of data taken in the final
2011 JVLA epoch.  We found that the CLEAN bias was not
stable and seemed to correlate with the amount of interference in
the data. In particular, there were three days when the CLEAN bias
was significantly higher than usual. The higher CLEAN bias on those
days appeared to result from CLEANing too deeply, i.e.,
since there was more interference, the rms was higher and hence using
the standard CLEAN limit was inappropriate. For those days the
images were remade with a shallower CLEAN threshold, which
brought the CLEAN bias back into line with the normal level.

The price of shallower CLEANing for those fields was an increase
in the rms noise levels.  Fig.~\ref{fig-gridrms}
shows the distribution of the rms noise levels in all \FIRST\ ``grid''
images (single pointings, before co-adding), with the rms distribution
for images on the three noisy JVLA days highlighted.  The mode of
the noise distribution is 0.132~mJy, but the distribution has a
tail toward higher rms values (most often the result of bright
sources in the field).  The extension to lower rms values is the
result of multiply observed fields that have been combined to reduce
the noise. The noise levels for JVLA fields observed on the three
bad days are typically 2.5 times higher.  Note that the \FIRST\ catalog
uses these noise measurements for all contributing grid images to
compute the rms noise as a function of position in the co-added
images \citep{white1997}; consequently, the final \FIRST\ catalog
has rms noise estimates that correctly reflect the elevated noise
in these regions.

\subsection{Epochs of Observations}

To enable the use of \FIRST\ data for time-domain science,
the 2014Dec17 release of the \FIRST\ catalog includes observation epochs
for the sources.  Since the catalog is created from co-added images, each
source may have contributions from many different pointings taken at different times.
For some sources all the contributing observations come from a narrow time range of
only a few minutes (when one or two consecutive grid images dominate the co-added
source position), while others have significant contributions from pointings taken
years apart (typically at the seams between observing seasons and near fields that
were re-observed owing to problems with the original observations).

The catalog includes a mean weighted epoch ${\bar t}$, defined to be the
average of the epochs of all the contributing pointings at the source position weighted
by the same weights used to combine the overlapping maps.  It also includes the
weighted rms $\sigma(t)$ of the scatter of the pointing epochs about that mean, which is
a measure of the effective spread in the observing epoch.  Figure~\ref{fig-rmsdist}(a)
shows the cumulative distribution of $\sigma(t)$; the median rms is 1.8 days, and 90\% of the $\sigma(t)$ values
are less than four months, so most observations
are well-localized in time.  Plotting $\sigma(t)$ versus ${\bar t}$
(Fig.~\ref{fig-rmsdist}b) shows the tail of much higher epoch rms values for
sources in the overlap regions between observing sessions.

\section{Astrometry}\label{astrometry}

 In our initial description of the \FIRST\
survey \citep{becker1995} we evaluated the astrometric precision
of the source catalog by comparing the positions of 46 Multi-Element Radio-Linked
Interferometer Network (MERLIN)
calibrators lying in the first survey strips and found the systematic
errors to be $<0.05\arcsec$. An additional comparison with 4100 optical
counterparts from the Automatic Plate Measuring (APM) machine scans of the Palomar
Optical Sky Survey plates \citep{mcmahon1992} found the
same limit for the optical-radio frame offset. For a survey in which
the individual catalog entries have typical positional uncertainties
of $0.3\arcsec$, any systematic position offsets of this magnitude
were insignificant for all envisioned initial usages of the catalog.
In the \FIRST\ catalog paper \citep{white1997} we used a larger
sample of radio calibrators and the increased size of the survey
coverage area to conclude that any offset from the radio reference
frame was $<0.03\arcsec$.

\begin{deluxetable*}{cclcccccc} 
\tabletypesize{\footnotesize} 
\tablecolumns{9}
\tablewidth{0pt}
\tablecaption{\FIRST\ Matches with External Catalogs \label{table-match}}
\tablehead{
\colhead{FIRST} & \colhead{Subset} & \colhead{External} & \colhead{Match} & \colhead{Match} &
\multicolumn{2}{c}{Position rms $\sigma$\tablenotemark{c}} & \multicolumn{2}{c}{Position mean offset\tablenotemark{d}} \\
\colhead{Subset} & \colhead{Percentage\tablenotemark{a}} & \colhead{Catalog} & \colhead{Radius} & \colhead{Percentage\tablenotemark{b}} &
\colhead{$\Delta\hbox{RA}$} & \colhead{$\Delta\hbox{Dec}$} & \colhead{$\Delta\hbox{RA}$} & \colhead{$\Delta\hbox{Dec}$} \\
\colhead{} & \colhead{(\%)} & \colhead{} & \colhead{(arcsec)} & \colhead{(\%)} &
\colhead{(arcsec)} & \colhead{(arcsec)} & \colhead{(arcsec)} & \colhead{(arcsec)} \\
\colhead{(1)} & \colhead{(2)} & \colhead{(3)} & \colhead{(4)} & \colhead{(5)} &
\colhead{(6)} & \colhead{(7)} & \colhead{(8)} & \colhead{(9)}
}
\startdata
all                         & 100.0\phn & NVSS           & 15\phn & 54.9    & 5.46 &  5.59 & $ {-}0.009\pm0.008$\phn & $   0.064\pm0.008 $\\
isolated\tablenotemark{e}   &  73.0     & NVSS           & 15\phn & 50.6    & 4.64 &  4.89 & $ {-}0.011\pm0.008$\phn & $   0.066\pm0.008 $\\[5pt]

all                         & 100.0\phn & 2MASS          & 2      & \phn8.0 & 0.42 &  0.43 & $    0.021\pm0.002$     & $   0.007\pm0.002 $\\
pt\tablenotemark{f}         &  37.1     & 2MASS          & 2      & \phn8.3 & 0.34 &  0.34 & $    0.020\pm0.002$     & $   0.006\pm0.002 $\\
pt, bright\tablenotemark{g} & \phn4.9   & 2MASS          & 2      & \phn6.8 & 0.23 &  0.23 & $    0.022\pm0.004$     & $   0.009\pm0.004 $\\[5pt]

all        &  \phn93.0\tablenotemark{h} & SDSS all       & 2      & 32.9    & 0.36 &  0.37 & $    0.021\pm0.001$     & $   0.016\pm0.001 $\\
all               &  93.0 & SDSS galaxy\tablenotemark{i} & 2      & 26.5    & 0.38 &  0.39 & $    0.021\pm0.001$     & $   0.016\pm0.001 $\\
all               &  93.0 & SDSS star\tablenotemark{j}   & 2      & \phn6.3 & 0.26 &  0.28 & $    0.023\pm0.001$     & $   0.014\pm0.001 $\\
pt                          &  35.3     & SDSS all       & 2      & 38.2    & 0.28 &  0.28 & $    0.020\pm0.001$     & $   0.014\pm0.001 $\\
pt                          &  35.3     & SDSS galaxy    & 2      & 29.6    & 0.30 &  0.30 & $    0.020\pm0.001$     & $   0.015\pm0.001 $\\
pt                          &  35.3     & SDSS star      & 2      & \phn8.6 & 0.21 &  0.22 & $    0.022\pm0.001$     & $   0.012\pm0.001 $\\
pt, bright                  & \phn4.7   & SDSS all       & 2      & 41.0    & 0.17 &  0.17 & $    0.022\pm0.001$     & $   0.012\pm0.001 $\\
pt, bright                  & \phn4.7   & SDSS galaxy    & 2      & 23.4    & 0.19 &  0.19 & $    0.020\pm0.002$     & $   0.013\pm0.002 $\\
pt, bright                  & \phn4.7   & SDSS star      & 2      & 17.7    & 0.15 &  0.16 & $    0.025\pm0.002$     & $   0.011\pm0.002 $\\[5pt]

JVLA\tablenotemark{k}, isolated & 4.5   & NVSS           & 15\phn & 52.0    & 4.62 &  4.99 & $    0.005\pm0.030$     & $   0.179\pm0.031 $\\
JVLA                        & \phn6.4   & 2MASS          & 2      & \phn6.7 & 0.47 &  0.49 & $    0.000\pm0.007$     & ${-}0.029\pm0.008\phn $\\
JVLA                        & 99.3      & SDSS all       & 2      & 28.4    & 0.40 &  0.44 & $    0.002\pm0.003$     & ${-}0.009\pm0.003\phn $\\
\enddata
\tablenotetext{a}{Percentage of \FIRST\ catalog sources included in the selected subset.}
\tablenotetext{b}{Percentage of the \FIRST\ subset that has matches in the external catalog.}
\tablenotetext{c}{RMS difference between \FIRST\ and external catalog positions. For NVSS the 68.3\% percentile width
of the distribution is reported because the NVSS difference distribution is highly non-Gaussian (Fig.~\ref{fig-NVSS_astrometry}).}
\tablenotetext{d}{Mean value of \FIRST\ position minus external catalog position.}
\tablenotetext{e}{Isolated \FIRST\ sources having no neighbors within 50\arcsec.}
\tablenotetext{f}{\FIRST\ point sources having fitted FWHM major axes $< 6\arcsec$.}
\tablenotetext{g}{Bright \FIRST\ sources having peak flux densities $> 10$~mJy.}
\tablenotetext{h}{Includes only sources in sky area covered by SDSS DR10.}
\tablenotetext{i}{SDSS sources classified as galaxies (type $=3$).}
\tablenotetext{j}{SDSS sources classified as stars (type $=6$).}
\tablenotetext{k}{Includes only sources from 2011 JVLA observations.}
\end{deluxetable*} 


With the survey now complete, large optical and infrared
catalogs of the sky now available, and significant work over the intervening
two decades invested in the creation of the International Celestial
Reference Frame \citep[ICRF2 --][]{ma2013}, we have the opportunity to
investigate small astrometric effects in some detail. In subsequent
sections, we provide definitive matches of the \FIRST\ survey's
final catalog to the NVSS, SDSS, and 2MASS catalogs; Table~\ref{table-match} summarizes
the results of those matches, including the mean astrometric offsets between
\FIRST\ and other catalogs. 

The positional offset for all \FIRST-NVSS matches in
Right Ascension is consistent with zero: $\Delta RA = -0.009 \pm
0.008\arcsec$ (avoiding source confusion by selecting isolated
sources only, the value is $\Delta RA = -0.011 \pm 0.008\arcsec$).
This agreement is unsurprising given that both surveys were conducted for the
most part with the same instrument, but it eliminates the possibility
that any errors have been introduced by the data reduction procedures.

Comparisons with the optical and near-IR catalogs, however, reveal
an apparent discrepancy. In Table~\ref{table-match},
the mean offset in RA between \FIRST\ and either SDSS or 2MASS is
consistent with $0.021\arcsec$, and the offset is determined with a very
small uncertainty ranging from 1 to 4 mas for the various subsets.
We note that an A-configuration
survey of the SDSS Stripe 82 region \citep{hodge2011} also finds
$\Delta RA = 0.020\arcsec$. Since the offset between the ICRF2 and
optical reference frames has been established to be less than
$0.003\arcsec$ in RA \citep{orosz2013,assafin2013},
it appears that VLA data suffer from a $+20$~mas offset.

However, using only \FIRST\ data taken with the JVLA and matching
to the SDSS catalog, we find $\Delta RA = 0.002\arcsec\pm 0.003$.
The timing system, correlator, and data acquisition hardware and
software are different between the VLA and the JVLA.
Thus, we find strong evidence
that positions derived from the old VLA system have a systematic
offset in RA of $\sim +20$~mas with respect to the radio and optical
reference frames.

Note that this systematic error is very much smaller than the 
positional uncertainties on sources in the \FIRST\ catalog;
for even the brightest sources, the random position errors are
$\sim100$~mas or larger.  This error is small enough to be irrelevant
except in large-scale cross-matches that compare astrometric
systems.

The declination offsets also show some systematic variations in
Table~\ref{table-match}, although different external catalogs
have different offsets.   The SDSS matches have a typical declination
offset of 15~mas, while the 2MASS offset is smaller, $\sim7$~mas.
The NVSS-\FIRST\ declination offset stands in stark contrast with a difference of 60~mas.
As for Right Ascension, in the JVLA region the remaining declination offsets
compared with SDSS
are consistent with zero.  
We discount the large discrepancies from the NVSS
match, which we find in \S\ref{NVSS} to have highly non-Gaussian
position difference distributions with long tails.  With that caveat, we conclude from the
SDSS and 2MASS matches
that systematic errors in the \FIRST\
declinations at all epochs are of order 0.02\arcsec\ or less.

\begin{figure}[b] 
\includegraphics[width=\linewidth]{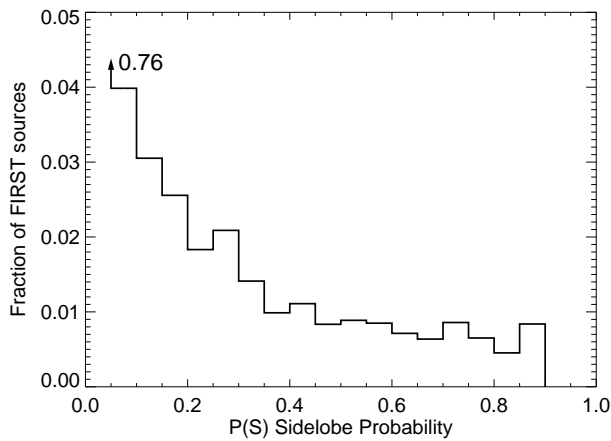}
\caption[Sidelobe probability distribution]{Distribution of sidelobe
probability $P(S)$ in the \FIRST\ catalog.  76\% of the sources
are in the lowest bin, $P(S) < 0.05$.
\label{fig-sidehist}}
\end{figure}

\begin{figure*}
\includegraphics[width=0.5\linewidth]{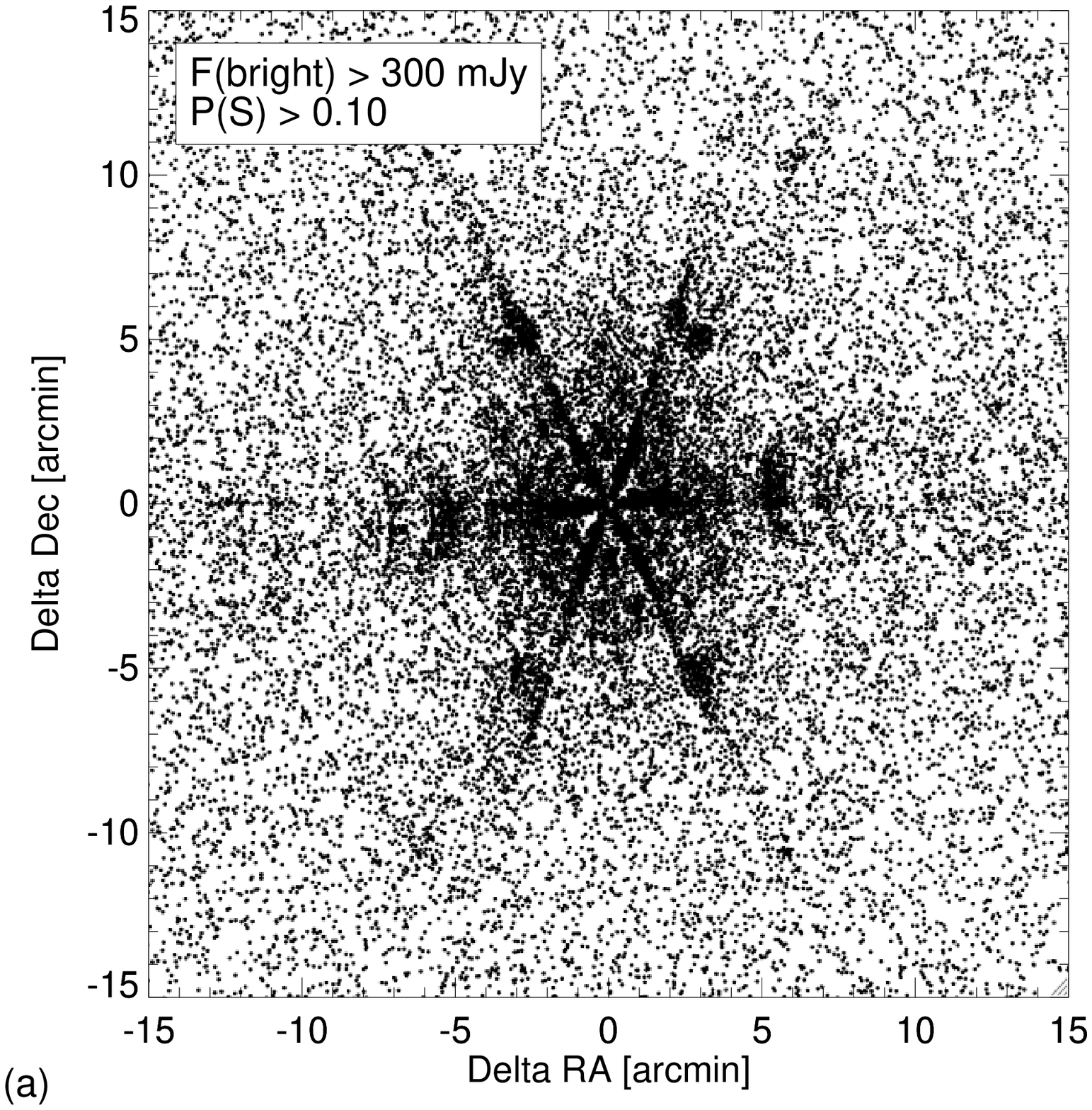}
\includegraphics[width=0.5\linewidth]{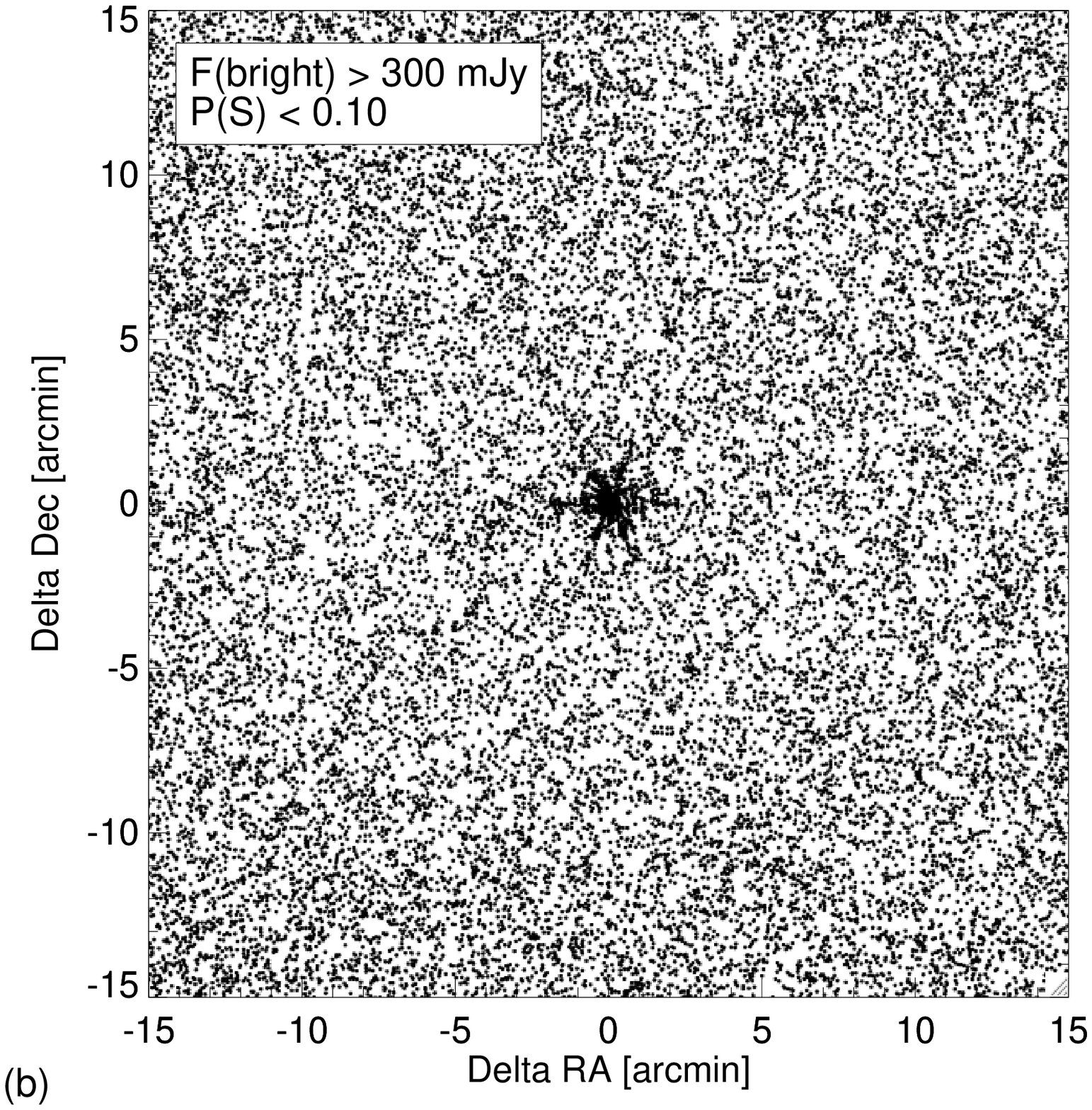}
\caption[Sidelobe distribution around bright sources]{Distribution of sidelobes
in the vicinity of bright sources.  The offsets of all \FIRST\ catalog entries 
in the vicinity of sources with $F(\hbox{peak}) > 300$~mJy are shown.  (a) The typical snapshot
sidelobe pattern that results from the three arms of the VLA is clearly visible
in sources with higher sidelobe probabilities, $P(S) > 0.1$.
(b) The strong sidelobe pattern is almost
absent for the sources with low sidelobe probabilities, $P(S) < 0.1$.
\label{fig-sidedist}}
\end{figure*}

\begin{figure}
\includegraphics[width=\linewidth]{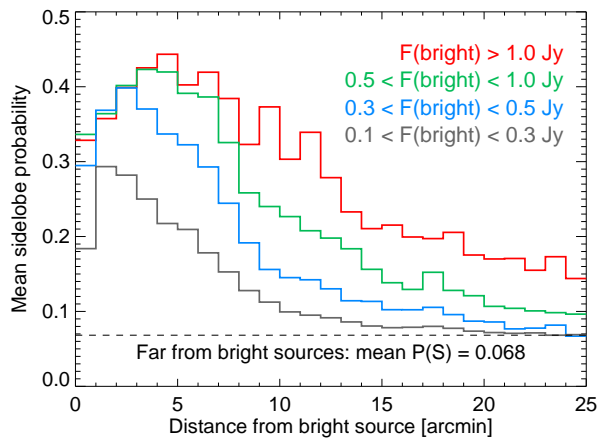}
\caption[Sidelobe radial dependence]{Variation of mean sidelobe
probability $P(S)$ with distance from bright sources.  The
sidelobe likelihood generally declines with the flux density
of the bright source and with distance from the bright source,
as expected.  The dashed line shows the mean $P(S)$ for
all the catalog sources that are not within 25~arcmin of
a source brighter than 100~mJy.
\label{fig-sideradial}}
\end{figure}

\begin{figure}
\includegraphics[width=\linewidth]{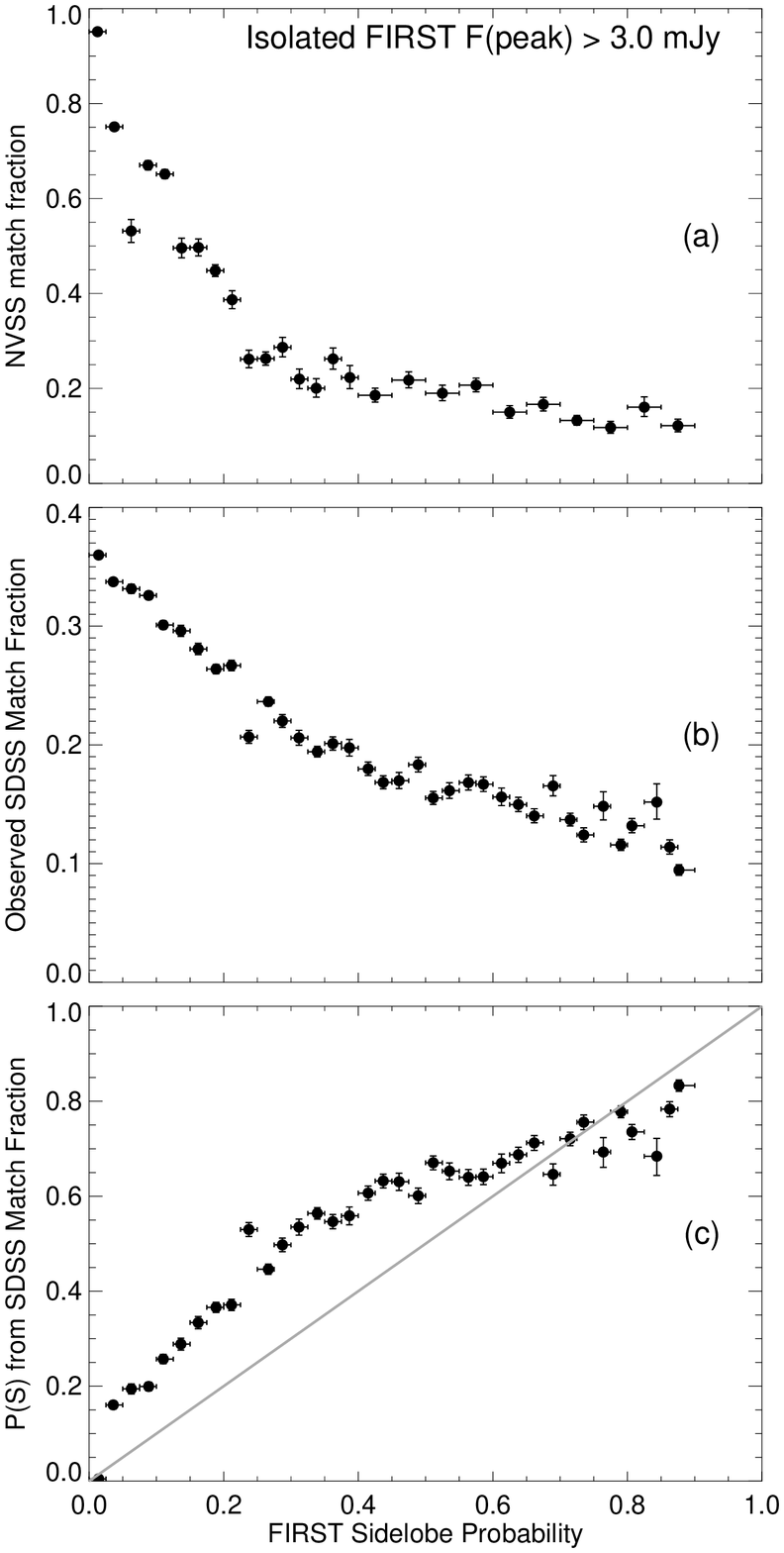} 
\caption[NVSS and SDSS sidelobe tests]{Tests of the accuracy
of the sidelobe
probability $P(S)$. (a) \FIRST-NVSS match rate as
a function of $P(S)$.  Only
isolated \FIRST\ sources with integrated flux densities
brighter than 3.0~mJy are included.
For low $P(S)$ values,
95\% of the \FIRST\ sources are detected in NVSS.
(b) \FIRST-SDSS match rate as
a function of $P(S)$.
The decline in the match rate with increasing sidelobe
probability is roughly consistent with expectations
assuming that the fraction of real sources in each
bin is proportional to $1-P(S)$.  The distribution is
not expected to be exactly linear since the radio flux density
distribution varies with $P(S)$, and the observed SDSS match rate depends on
radio flux (Fig.~\ref{fig-2MASS-SDSSfractions}).
(c) Sidelobe probability esimated using the SDSS match rates in (b).
The effect of variation of match rate with radio flux was corrected
using a fit to the distribution in Fig.~\ref{fig-2MASS-SDSSfractions}.
The points would lie along the diagonal line if the catalog $P(S)$
values were perfect predictors of the sidelobe fraction.
The sidelobe probability $P(S)$ quoted in the catalog ($x$-axis) appears
to be slightly underestimated for moderate probabilities
($0.1 < P(S) < 0.6$).
\label{fig-sidelobes}}
\end{figure}

\section{The Sidelobe Flagging Algorithm}\label{sidelobes}

The original versions of the \FIRST\ catalog (beginning with the 
1995 October 16 release) included a sidelobe warning flag that
indicated a likelihood that the source entry was actually
a sidelobe of a bright nearby source rather than a real
object \citep{white1997}.  The algorithm for setting this
flag used an oblique decision tree
classifier\footnote{The OC1 oblique decision tree software is
available for download at
\url{http://www.cbcb.umd.edu/\textasciitilde{}salzberg/announce-oc1.html}.}
\citep{murthy1994}
that was trained using a set of sidelobes identified by visual
examination of some \FIRST\ images.

While this approach had some value in identifying potentially
spurious sources, it was not very accurate.  It was easy to find
cases where sidelobes were not flagged or real sources were
incorrectly flagged.  In addition, the use of a binary yes/no
flag for sidelobe flagging did not provide much guidance as
to the actual likelihood that a source was spurious.  Consequently,
this catalog entry was considered somewhat unreliable and saw relatively little use.

\cite{white2005} developed a more sophisticated and
useful variation on this algorithm.  We used deep observations of
a portion of the Galactic plane from a different survey to determine objectively which
sources in a catalog were spurious (not
seen in the much deeper data) and which were real (confirmed in
the deeper data).  That produced a reliable training set for
the decision tree classifier.  We then created multiple
independent decision trees whose output was combined to obtain
a sidelobe probability estimate for each source rather than a simple
binary classification.  This voting decision tree approach is
described in more detail by \cite{white2000} and \cite{white2008}.

For the 2008 and later releases of the \FIRST\ catalog, we adopted a
similar approach to computing sidelobe probabilities for the \FIRST\
sources.  We created a training set of 1905 sources (including 120
sidelobes).  We matched \FIRST\ with two
deep radio surveys, the Spitzer First Look Survey
\citep[416 \FIRST\ sources, 15 sidelobes;][]{condon2003}
and the COSMOS deep
survey pilot area \citep[72 \FIRST\ sources, 6 sidelobes;][]{schinnerer2004}\footnote{The
deeper imaging from \cite{schinnerer2007} was not available at the time.}.
We also included a less reliable but larger sample from our own
survey of the Deeprange area \citep[1356 \FIRST\ sources,
99 sidelobes;][]{white2003}. Finally we augmented the set with a sample of 61
bright sources ($F>2$~Jy), none of which are sidelobes; this
improved the performance of the classifier for bright
objects.

The 15 parameters used for the classification include nine source
properties (peak-to-integrated flux ratio, rms noise level, source
major and minor axes compared with the synthesized beam, source position angle,
ratio of source peak and integrated flux densities to the corresponding
values in the ``island'' to which the source belongs\footnote{The islands are rectangular pixel regions
surrounding sources as described in \S3 of \cite{white1997}.}), and six
properties of the nearest bright source that could be creating
sidelobes (positional offsets, flux ratios, and directions). Ten
independent oblique decision trees were created using this training
set.  Their outputs are combined as described in \cite{white2000}
to estimate a sidelobe probability, $P(S)$, for every \FIRST\ source.

A histogram of the resulting sidelobe probabilities is shown in
Figure~\ref{fig-sidehist}.  The vast majority of the sources have
low sidelobe probabilities; in fact, 70\% of the objects have the
minimum $P(S)$ value of 0.014.  The mean sidelobe probability for
the catalog is 0.097, which is an estimate of the fraction of sidelobes
in the catalog.

Figure~\ref{fig-sidedist} shows the spatial distribution
of objects in the vicinity of bright radio sources, which frequently
give rise to nearby sidelobes.  Applying a probability cut $P(S) < 0.1$
(Fig.~\ref{fig-sidedist}b) eliminates the vast majority of the obvious
sidelobe pattern.  The effect of the bright source can also be
seen in the radial dependence of the mean $P(S)$ (Fig.~\ref{fig-sideradial}).
The probability is lower in the closest bins because of the tendency of
bright radio sources to have multiple components (doubles, triples, etc.);
away from the center the probability declines with radius and with the
bright source flux density.

The sidelobes are heavily concentrated in the vicinity of bright sources.
The 10\% of \FIRST\ catalog sources that fall within 10~arcmin of a bright
object ($F > 100$~mJy) have a mean sidelobe fraction $P(S) = 0.24$ and account
for 25\% of all the sidelobes in the catalog.  The remaining 90\% of \FIRST\ sources (not
near a bright object) have a much lower sidelobe fraction, $P(S) = 0.08$.
We conclude that the sidelobe probability behaves qualitatively as
expected, and that the $P(S)$ value does separate likely sidelobes
from other sources in the vicinity of bright objects.

We rely on matches to external catalogs for quantitative assessments
of the accuracy of the \FIRST\ sidelobe probabilities.  The
NVSS-\FIRST\ cross-match (\S\ref{NVSS}) provides one obvious
test of $P(S)$.  It is not definitive, however, because the
2.5~mJy NVSS detection limit is significantly shallower than \FIRST.
The great majority of sources with high
$P(S)$ values are faint in the radio: the mean peak flux density
for sources with $P(S)>0.1$ is only 1.35~mJy, and only $\sim4$\%
of those objects have flux densities above the NVSS detection
threshold. That means that NVSS can only be used to confirm the
accuracy of $P(S)$ for bright sources; it does not provide any
information about the much more common faint sidelobes.  Nonetheless,
the comparison is useful.  Figure~\ref{fig-sidelobes}(a) shows the NVSS
detection fraction for \FIRST\ sources as a function of sidelobe
probability.  We include only \FIRST\ sources that are bright
enough that they ought to be detected by NVSS.  The detection
fraction declines as expected as $P(S)$ increases.
The general trend toward fewer detections at higher
$P(S)$ is clear and confirms that the sidelobe probabilities are
reliable for brighter \FIRST\ sources.

The cross-match between \FIRST\ and SDSS (\S\ref{SDSS}) provides
a more powerful test of the sidelobe probabilities.  That may be
surprising since the optical counterparts of most \FIRST\ sources 
are too faint to be detected by SDSS;
the absence of an optical counterpart does not reveal much about
the reality of an individual radio source.  However, the SDSS match
fraction provides an accurate \textit{statistical} measurement of
the sidelobe fraction down to the detection limit of the \FIRST\
survey.  The complication in this case is that the fraction of
sources with optical counterparts depends on the radio flux density,
but that variation is relatively smooth and can be easily modeled.

Figure~\ref{fig-sidelobes}(b) displays the fraction of \FIRST\
sources with an SDSS counterpart as a function of the sidelobe
probability.  The distribution is well-behaved and clearly shows
the expected decline with the increasing fraction of spurious sources
(which naturally do not have SDSS counterparts).  In this figure
no corrections have been made for the dependence of the SDSS match
fraction on radio flux density.  For Figure~\ref{fig-sidelobes}(c),
we have both corrected for that flux dependence and also have inverted
the distribution, using the corrected SDSS match fraction as a
measurement of the sidelobe probability.  This can then be compared
directly with the catalog value for $P(S)$.  The points would fall
along the diagonal line if the catalog $P(S)$ values were perfect
predictors of the sidelobe fraction.  The actual sidelobe probabilities
appear to be slightly higher than the catalog estimates, but the
catalog $P(S)$ values do appear to be reasonably accurate.

In summary, the sidelobe probabilities $P(S)$ in the \FIRST\
catalog give a useful measure of the likelihood that any source
is spurious.  The probabilities have been shown to be sufficiently
accurate to be useful, and they do a good job of eliminating the spurious detections that
tend to cluster around bright radio sources.

\begin{figure}
\includegraphics[width=\linewidth]{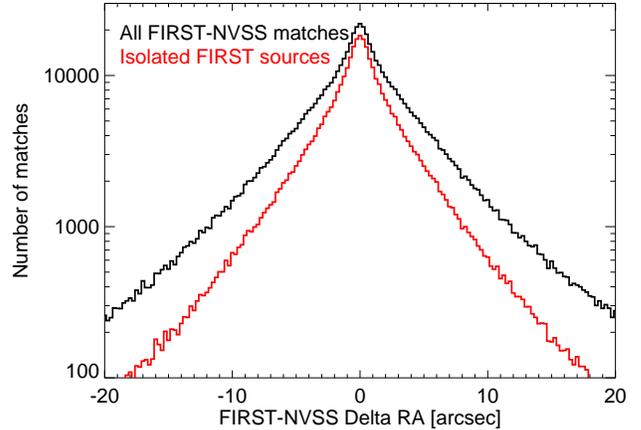}
\caption{The positional offsets in RA for all \FIRST-NVSS radio
source matches (in black) and for isolated \FIRST\ sources
defined as there being only one such source within 50\arcsec\
(the NVSS synthesized beam FWHM) of an NVSS source. The distribution
is highly symmetrical and centered very close to zero (see Table~\ref{table-match}
for quantitative details). The distribution for all sources is
significantly wider as single NVSS sources are often resolved into
multiple components by \FIRST.\label{fig-NVSS_astrometry}}
\end{figure}

\begin{figure*}
\includegraphics[width=0.5\linewidth]{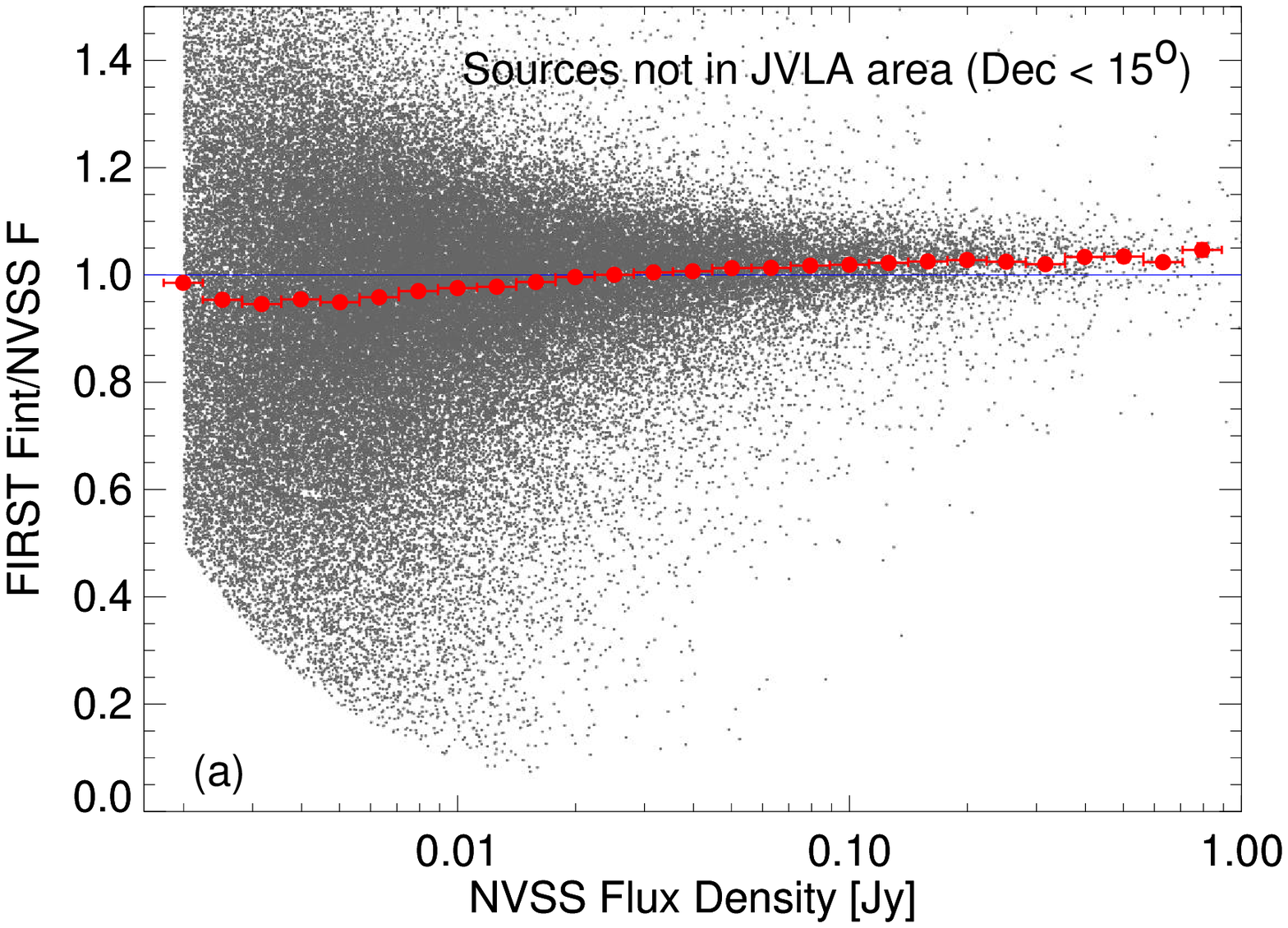} 
\includegraphics[width=0.5\linewidth]{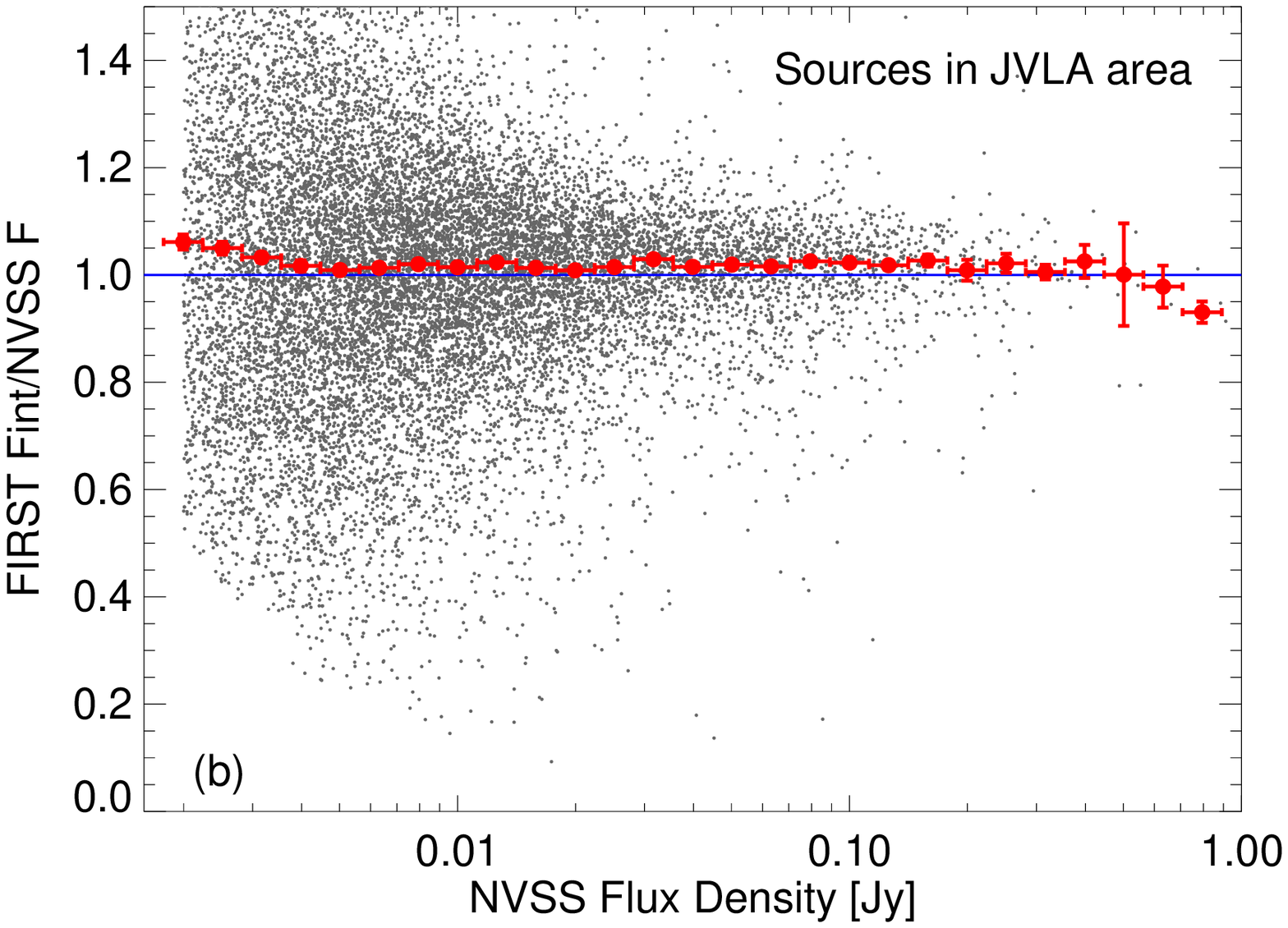} 
\caption{The ratio of \FIRST\ source flux density to NVSS flux
density as a function of NVSS flux density
(a) for all matching sources at declinations $<15^{\circ}$
and
(b) for sources observed in 2011 exclusively with the JVLA.
The brighter of the peak or
integrated \FIRST\ flux density is used. The blue bar shows a
ratio of unity, while the red dots show the median values for the
bins indicated by the horizontal error bars (vertical error bars
are mainly smaller than the points). The 2.0~mJy cutoff on the left is the
NVSS threshold; the curving cutoff in the lower left represents the
\FIRST\ threshold of 1.0~mJy.\label{fig-NVSS_FIRST_ratio}}
\end{figure*}

\begin{figure}[b] 
\includegraphics[width=\linewidth]{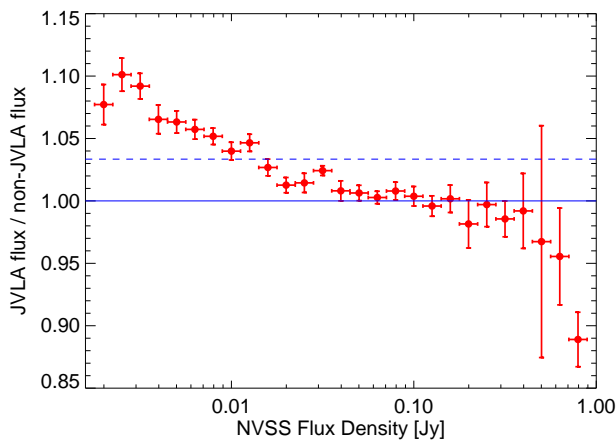}
\caption{The ratio of the flux density offsets (\FIRST\ divided
by NVSS) between \FIRST\ data taken with the VLA and the JVLA.
All data are for sources with $\delta < 15^{\circ}$ to eliminate
declination effects. The horizontal dashed line at $+3.3\%$ represents the
expected offset for sources with a mean spectral index of
$\alpha = -0.7$, given the difference in effective frequency for
the two set of observations.\label{fig-nonJVLA_JVLA_ratio}} \end{figure}

\begin{figure}[b] 
\includegraphics[width=\linewidth]{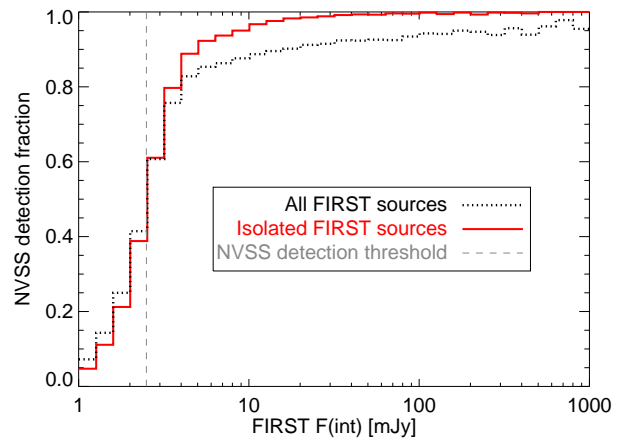}
\caption[NVSS detection fraction]{Fraction of
\FIRST\ sources that are detected by NVSS
as a function of the \FIRST\ integrated flux
density using a 15\arcsec\ matching radius.
The vertical dashed line shows the NVSS 2.5~mJy detection limit.
The black dotted line shows the distribution for all \FIRST\ sources.
The small fraction of bright unmatched sources
are extended objects that are
resolved into multiple components by \FIRST.  That is demonstrated clearly
by the solid red line, which
shows the distribution for isolated \FIRST\ sources (having no neighbors
within 50\arcsec).
\label{fig-nvssdet}}
\end{figure}

\section{Comparisons with NVSS: Completeness and Reliability}\label{NVSS}

The NRAO VLA Sky Survey \citep[NVSS ---][]{condon1998}, also conducted
at 20~cm, covered over $3\pi$ steradians of the sky north of $\delta =
- 40^{\circ}$ to a completeness limit outside of the Galactic plane
of roughly 2.5~mJy. The synthesized beam size was $\sim 45\arcsec$
leading to rms positional uncertainties of 7\arcsec\
for point sources at the catalog detection limit, with errors for brighter sources decreasing
inversely with flux density
to $\sim 1$\arcsec\ for the brightest sources\footnote{But see \S\ref{resolution} for
a discussion of more realistic positional uncertainties for optical matching.}. The large beam size
allowed the detection of extended, low-surface-brightness objects
that can be resolved out by the high-resolution 
(5\arcsec) \FIRST\ beam; in addition, for sources with angular sizes
between $\sim 10\arcsec$ and 60\arcsec, the 
\FIRST\ flux densities underestimate the true integrated source
intensity.

In Figure~\ref{fig-NVSS_astrometry} we show the positional offsets in
RA for all \FIRST-NVSS radio source matches (in black) and for
all isolated \FIRST\ sources defined such that there is only
one such source within 50\arcsec\ of the NVSS source
position (roughly equal to the NVSS synthesized beam FWHM). The distribution is
highly symmetrical and centered very close to zero (see Table~\ref{table-match} for
quantitative details); the distribution of offsets in declination
is indistinguishable from the RA distribution. The distribution for all sources is significantly
wider as single NVSS sources are often resolved into multiple
components by \FIRST. Both distributions are distinctly
non-Gaussian. Thus, rather than quote an rms in Table~\ref{table-match}, we record
the 68.3\% percentile of the absolute value distribution, which
is the equivalent of the rms for a Gaussian distribution.

In Figure~\ref{fig-NVSS_FIRST_ratio}(a) we plot the ratio of \FIRST\
source flux density to NVSS flux density for all matching sources
at declinations $<15^{\circ}$ (imposed for comparison with
Fig.~\ref{fig-NVSS_FIRST_ratio}b) as a function of NVSS flux
density. The brighter of the peak or integrated \FIRST\ flux
density is used. The blue bar shows a ratio of unity, while the red
dots show the median values for the bins indicated by the horizontal
error bars (vertical error bars are smaller than the points). The
2.0~mJy cutoff on the left is the NVSS threshold; the curving cutoff
in the lower left represents the \FIRST\ threshold of 1.0~mJy.
The $\sim 1-5\%$ deficit in \FIRST\ flux density between 2.0 and
20~mJy is likely the result of diffuse flux from extended sources
resolved out by the \FIRST\ beam; the turn-up for the lowest
two flux density points arises from the imposition of the 1~mJy
\FIRST\ threshold that biases the distribution upward. The rise
above unity at NVSS flux densities $>50$~mJy could arise from a
calibration offset of $\sim 2\%$ plus a (possibly dominant) contribution from the
different bandwidths used in the two surveys convolved with the
source spectral index distribution.

Figure~\ref{fig-NVSS_FIRST_ratio}(b) displays the same plot as
Figure~\ref{fig-NVSS_FIRST_ratio}(a) for sources observed with the
JVLA in 2011. For the JVLA data, the agreement with NVSS fluxes is
actually somewhat better for flux densities greater than 4~mJy.  In
Figure~\ref{fig-nonJVLA_JVLA_ratio}, we plot the ratio of the flux
density offsets (\FIRST\ divided by NVSS) between \FIRST\
data taken with the VLA and with the JVLA. All data are for sources
with $\delta < 15^{\circ}$ to eliminate any declination-dependent
effects such as changes in the PSF.  We believe there are several
competing effects that produce the variations seen here.  First, sources
with typical spectral indices should be slightly brighter in the
lower-frequency JVLA observations (1.335 vs.\ 1.400~GHz -- see \S\ref{hardware}).
The horizontal dashed line at
$+3.3\%$ represents the expected offset for sources with a mean
spectral index of $\alpha = -0.7$.  That effect changes with flux,
however, since the spectral indices become flatter for compact
bright sources.  A second effect is that the lower-frequency JVLA
observations are also slightly lower resolution, which increases
their sensitivity to extended emission.  That is likely the reason
why the dip seen from 3--10~mJy in Figure~\ref{fig-NVSS_FIRST_ratio}(a)
is not seen in Figure~\ref{fig-NVSS_FIRST_ratio}(b).

This effect illustrates the sensitivity of
large surveys to small changes in observing parameters.
Subtle changes can lead to noticeable differences in large
statistical samples of sources, so flux differences between
the JVLA and VLA data at the level
of 5--10\% (Fig.~\ref{fig-nonJVLA_JVLA_ratio}) should be
treated with caution.  Correcting for these effects would require
knowledge of the spectral indices and sizes of the sources,
which are usually not available. Note, however, that these
systematic effects are small compared with the noise
for faint sources, so most studies can treat the JVLA and VLA
data as having equivalent flux scales. 

Figure~\ref{fig-nvssdet} shows the fraction of \FIRST\ sources
detected by NVSS as a function of the \FIRST\ integrated
flux density.  The factor of 10 difference between the resolutions
of the two surveys complicates the interpretation of this figure.
Many NVSS sources are resolved by \FIRST\ into multiple components.
Those sources will appear as lower flux \FIRST\ sources detected by NVSS,
when in fact NVSS only detects the sum of the components; that accounts
for most of the tail of detections at fainter fluxes.  Moreover,
such sources will often have large positional differences so that
they do not match within the 15\arcsec\ matching radius we are using
here; that accounts for the small fraction of bright \FIRST\ sources that
are undetected by NVSS.  The detection fraction for isolated \FIRST\
sources is also shown in Figure~\ref{fig-nvssdet}; essentially all bright
isolated \FIRST\ sources are detected by NVSS.

\begin{figure}
\includegraphics[width=\linewidth]{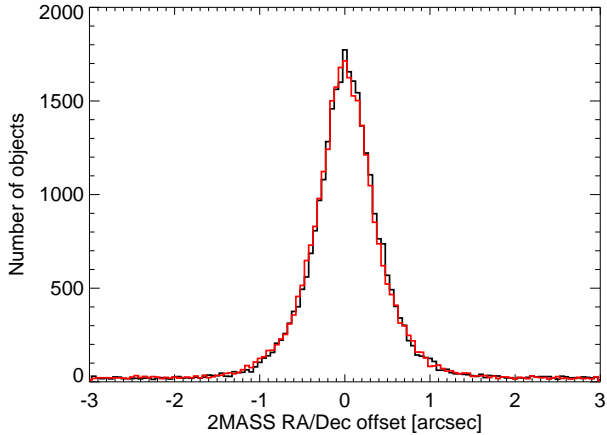}
\caption{The positional offsets in RA (black), and Dec (red) between
\FIRST\ radio sources and 2MASS objects. The distribution is
highly symmetrical and centered very close to zero (see Table~\ref{table-match} for
quantitative details).\label{fig-2MASS_astrometry}}
\end{figure}

\begin{figure}
\centering
\includegraphics[width=\linewidth]{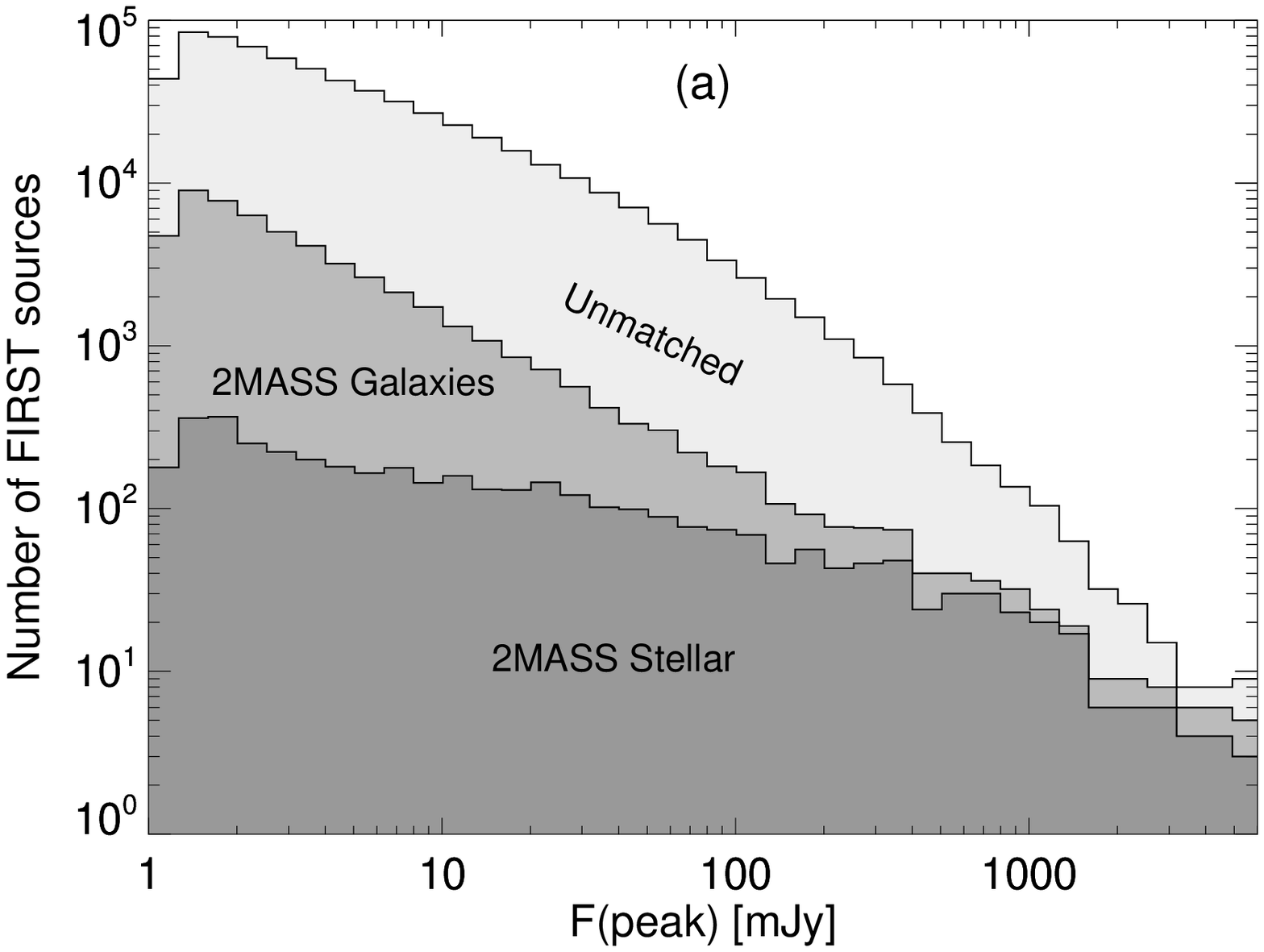} 
\includegraphics[width=\linewidth]{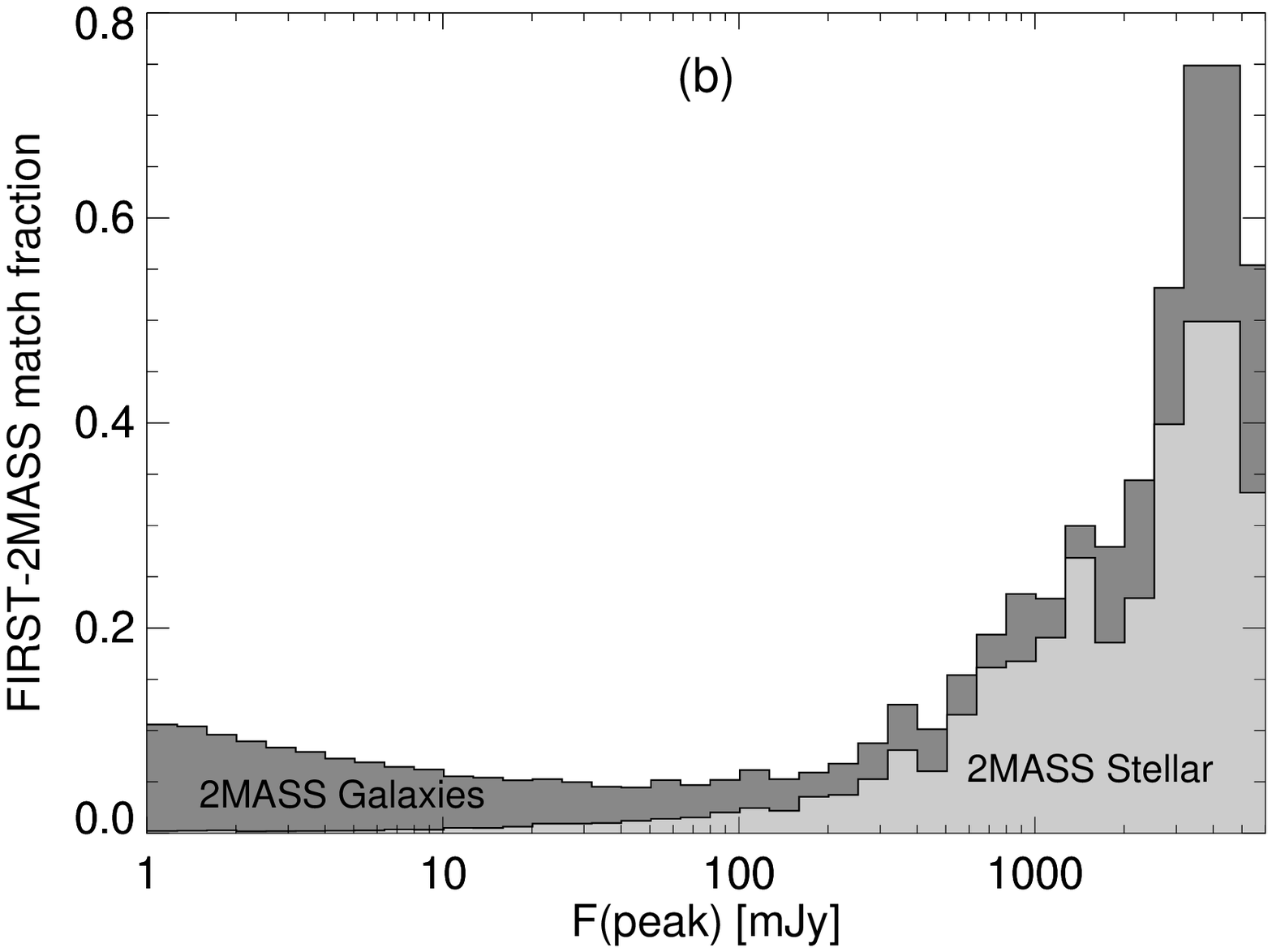} 
\caption{
\FIRST-2MASS matches as a function of \FIRST\ flux density.
The 2MASS sources have been divided into stellar objects (mainly AGN)
and galaxies using the SDSS classification.
(a) The raw counts of \FIRST\ sources and 2MASS matches
as a function of flux.
Only sources with low sidelobe probabilities, $P(S) \le 0.02$, are included,
which is why the number of sources drops near the detection limit..
(b) The fraction of 2MASS matches
as a function of flux.  The bulk of 2MASS sources detected by
\FIRST\ are galaxies.
\label{fig-2MASS_star_galaxy_fractions}}
\end{figure}

\section{Match to the 2MASS Catalog}\label{2MASS}

The first deep image of the entire sky at 2$\mu$m was produced by
the 2MASS Survey between 1997 and 2001 \citep{skrutskie2006}. The
primary data products from the survey are an image atlas, and point
and extended source catalogs containing over 470 million objects.

Astrometric data from the match of the \FIRST\ catalog to the
2MASS point source catalog are reported in Table~\ref{table-match} and displayed in
Figure~\ref{fig-2MASS_astrometry}.
We calculate the offsets between all \FIRST\ source
positions and any 2MASS object within $\pm 3\arcsec$ in each
coordinate; this large box
includes some chance coincidences but is highly
complete even for extended objects.
The astrometric offsets and rms widths are determined from Gaussian
fits to the distribution in each coordinate.
There is a statistically significant
$+20$~mas offset in Right Ascension and a $+10$~mas offset in
declination that persists even when only bright ($>10$~mJy) radio
sources are used in the matching; as noted in \S\ref{astrometry},
we believe that the RA offset, at least, arises from a systematic error in the VLA data
acquisition system which the JVLA has corrected.

The rms uncertainties are
the same in both coordinates. For the \FIRST\ convolving beam
size of 5.4\arcsec, even bright point sources ($>50 \sigma$)
will have an inherent rms positional uncertainty of $\sim
0.1\arcsec$ \citep{white1997}; the 2MASS 2.0\arcsec\ pixel size
convolved with variable seeing led to an astrometric accuracy of
$\lesssim 0.10\arcsec$ relative to the Hipparcos reference frame
for objects with $K_s<14$ \citep{skrutskie2006}. The fact that
the large majority of the 2MASS matches are extended galaxies (see
Fig.~\ref{fig-2MASS_star_galaxy_fractions})
plausibly makes up the remainder of the 0.23\arcsec\ value
reported in Table~\ref{table-match}.

\begin{figure}
\includegraphics[width=\linewidth]{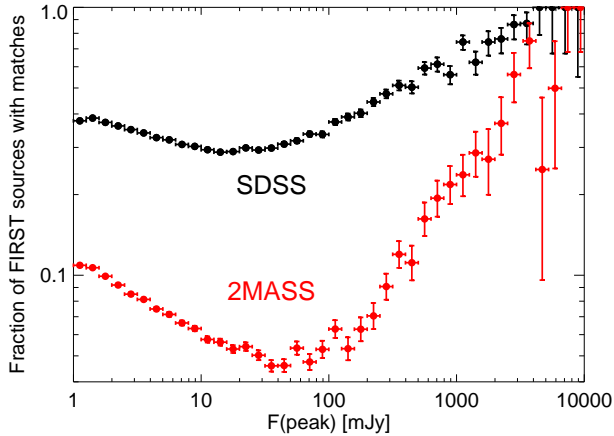}
\caption{The fraction of \FIRST\ radio sources with counterparts
in the SDSS and 2MASS catalogs (using a 2~arcsec matching radius) as a function of peak radio flux
density. Horizontal error bars represent the flux density ranges,
while vertical error bars represent statistical uncertainties.
Values are corrected for the false match rate.
Only sources with low sidelobe probabilities, $P(S) \le 0.02$, are included.
Both curves show a minimum where
the radio source counts change from being dominated by AGN at high
flux densities to being dominated by normal
galaxies.\label{fig-2MASS-SDSSfractions}}
\end{figure}

In Figures~\ref{fig-2MASS_star_galaxy_fractions}
and~\ref{fig-2MASS-SDSSfractions}, we show the fraction of
\FIRST\ radio sources with 2MASS counterparts as a function of radio
flux density. The match fractions have been corrected for the effects
of false matches using the density of 2MASS objects that fall
between 7.5\arcsec\ and 8\arcsec\ from the \FIRST\ source.
The \FIRST-2MASS match fraction falls from 100\% at the brightest flux
densities ($>5$~Jy) to less than $5\%$ at 40~mJy and then begins a steady
rise to 10\% near the survey threshold of 1~mJy. This is a consequence
of the two population components that comprise the radio $\log N-\log S$ curve in the 1--1000~mJy
range: ``monsters'' (radio-loud active galactic nuclei) and ``normal''
galaxies \citep{condon1992}; see the
end of \S\ref{SDSS} for further discussion of the effects of
the transition between populations.
This plot includes
only \FIRST\ objects with low sidelobe probabilities, $P(S) \le 0.02$ (\S\ref{sidelobes});
that avoids contaminating the counts
in the lowest flux bins with spurious sources.
The fact that the 2MASS points all lie below the
SDSS fraction of detections (Fig.~\ref{fig-2MASS-SDSSfractions}) is simply a consequence of the shallower depth of the 2MASS images.

\section{Match to the SDSS Catalog}\label{SDSS}

In recognition of the high scientific value to a radio survey of
having complementary optical data from which to derive radio source
identifications, the original \FIRST\ survey footprint was
designed to largely overlap the (then-planned) Sloan Digital Sky
Survey (SDSS). In the end, 93\% of the \FIRST\ sky coverage is
also covered by SDSS.  The extension to the original \FIRST\
footprint, approved in 2008, was added in order to provide complementary
data to a portion of the SDSS III survey.

The SDSS I and II projects collected imaging data between 2000 and
2008 over more than 10,000 deg$^2$ of the northern and southern
Galactic caps in five colors, as well as obtaining spectra of over
one million objects within the survey area. The images and catalogs
resulting from this effort are summarized in Data Release 7 \citep{abazajian2009}.
Data Release 10 in July 2013 increased the total
sky coverage to over 14,500 deg$^2$ and brought the catalog to
over one billion objects of which over 1.6 million have spectra
\citep{ahn2014}.  The \FIRST\ survey covers 68\% of the SDSS
DR10 sky area.

\begin{figure}
\includegraphics[width=\linewidth]{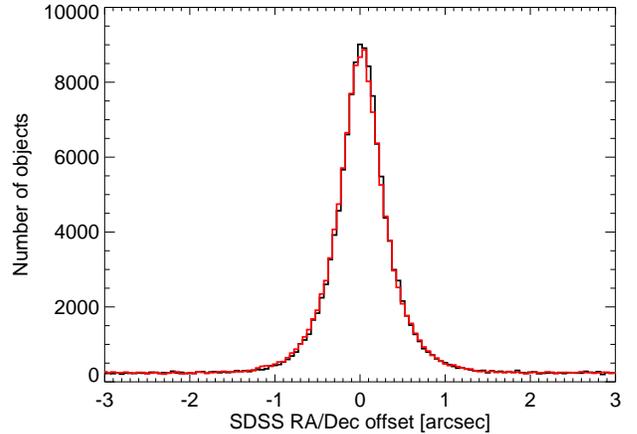}
\caption{The positional offsets in RA (black), and Dec (red) between
\FIRST\ radio sources and SDSS objects. As for the 2MASS matches, the distribution is
highly symmetrical and centered close to zero (see Table~\ref{table-match} for
quantitative details).\label{fig-SDSS_astrometry}}
\end{figure}

The first attempt at large-scale matching of SDSS and \FIRST\
sources was published by  \cite{ivezic2002}. The $\sim
0.1\arcsec$ astrometric offset in declination they found
was subsequently corrected in the SDSS astrometry pipeline.
Figure~\ref{fig-SDSS_astrometry} and Table~\ref{table-match} show the results of the
final match presented here for various optical and radio source
properties.  As with the 2MASS matches, we determine the astrometric
offsets and rms widths by fitting a Gaussian to the distribution
of SDSS sources found within
$\pm 3\arcsec$ of a \FIRST\ source in each
coordinate.
We report the match properties between all \FIRST\ and SDSS objects
as well as for subsets of the radio sources (point sources,
bright $>10$~mJy sources) matched with optical objects 
separated into stellar and galaxy counterparts.
As with 2MASS, we see a
small but statistically significant $+20$~mas offset in Right
Ascension and a $+10$ to~$+15$~mas offset in declination (\S\ref{astrometry}).

The uncertainties in RA and declination are essentially identical. For
the brighter ($S/N \gtrsim 50$) radio sources matched to point-like
optical counterparts, the rms of $\sim 0.15\arcsec$ is
consistent with the reported rms positional uncertainty for SDSS
($\sim 0.1\arcsec$), coupled with a similar uncertainty for
\FIRST\ sources.

\begin{figure}
\includegraphics[width=\linewidth]{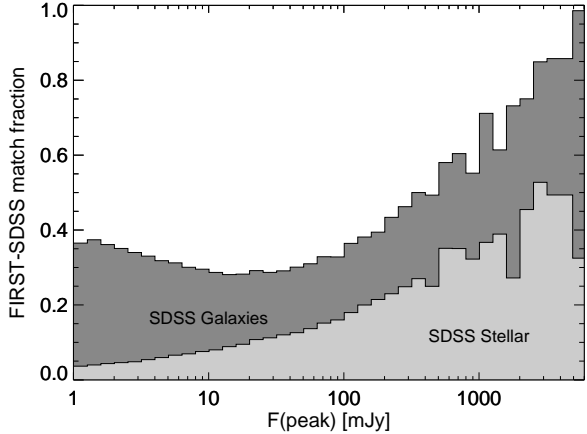}
\caption{The fraction of SDSS matches to \FIRST\ sources
as a function of \FIRST\ flux density using a matching
radius of 2~arcsec.
The SDSS sources have been divided into stellar objects (AGN)
and galaxies.
Only sources with low sidelobe probabilities, $P(S) \le 0.02$, are included.
\label{fig-SDSS_star_galaxy_fractions_linear}}
\end{figure}

\cite{ivezic2002} reported that roughly 30\% of the initial batch
of $10^5$ \FIRST\ sources had optical counterparts in SDSS, the
large majority of which (83\%) were resolved optical objects (i.e.,
galaxies); galaxies accounted for $\sim 50\%$ of the counterparts
at the brightest radio flux densities, rising to 90\% at the survey
threshold of $\sim 1$~mJy. The total fraction of SDSS matches in
the completed catalogs, shown in Figure~\ref{fig-SDSS_star_galaxy_fractions_linear},
is roughly consistent with this overall match rate.

\begin{figure}
\includegraphics[width=\linewidth]{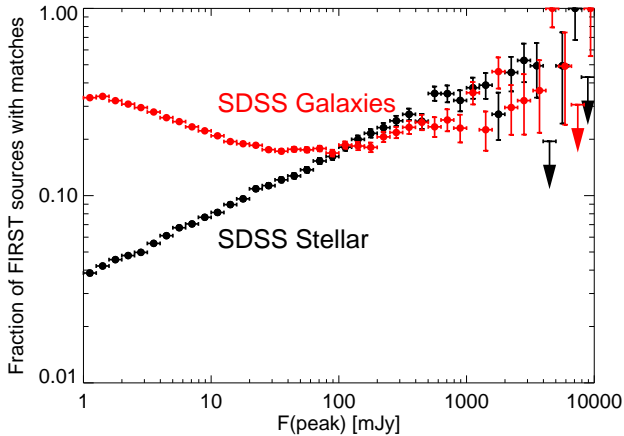}
\caption{The fraction of \FIRST\ radio sources with counterparts
in the SDSS divided by counterpart classification: stellar objects
which, with the exception of a handful of radio stars, are quasars
or other AGN, and galaxies.
Only sources with low sidelobe probabilities, $P(S) \le 0.02$, are included.
The fraction of sources with stellar counterparts
declines monotonically from 100\% to 3\%, while those with galaxy
counterparts transition from AGN-dominated systems at high flux
densities to radio emission dominated by star formation near the
survey threshold.
\label{fig-SDSS_star_galaxy_fractions}}
\end{figure}

Figures \ref{fig-SDSS_star_galaxy_fractions_linear} and
\ref{fig-SDSS_star_galaxy_fractions} show the SDSS match to the
completed \FIRST\ survey over a radio flux density range of a
factor of $10^4$. At the bright end, the majority of radio sources
have SDSS counterparts\footnote{In fact, just six of the 34 sources
brighter than 3~Jy fail to match an SDSS object within 1\arcsec. Three of these are
components of M87 which, of course, does have a match; one
falls outside the SDSS coverage; and the
other two are 3C280 and PKS~2127${+}$04, both
radio galaxies at $z=1$ that do fall below
the SDSS threshold.}. Stellar counterparts (all but a handful of
which are quasars) fall monotonically to $\sim 4\%$ of all
\FIRST\ sources at the survey threshold, while galaxy counterparts
fall to a minimum of a 10\% identification rate at 100~mJy and then
rise again to $>30$\% of all \FIRST\ sources at 1~mJy; this reflects
the long-established change in the radio source population mix as
a function of flux density \citep{condon1992}.

\begin{figure}
\includegraphics[width=\linewidth]{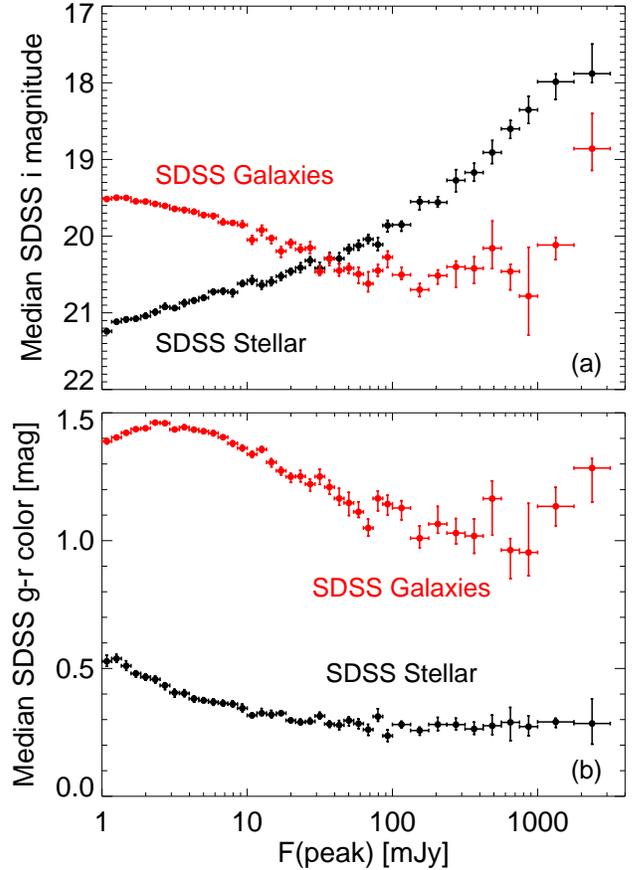} 
\caption{Magnitudes and colors of SDSS stars (black) and galaxies (red)
as a function of \FIRST\ peak flux density.
(a) Median $i$-band magnitudes.  Stellar objects
(mainly AGN) have radio and optical fluxes that decline in concert;
the flattening around $i=21$ occurs as the SDSS detection limit is
approached.  Galaxies, on the other hand, optically brighten
with decreasing radio flux density
as the population shifts
from high-redshift AGN in radio-bright galaxies to low-redshift
star formation associated with mJy sources.
(b) Median $g-r$ colors.  Stellar objects (quasars) are
typically more than 1 magnitude bluer than galaxies.  Both quasars and
galaxies get redder as the radio flux density decreases, presumably due to
the increasing redshift of fainter radio sources.  Galaxies begin to
get bluer for $F<3$~mJy as the nearby star-forming galaxies come into
view.
\label{fig-SDSS_mags_colors}}
\end{figure}

The magnitudes and colors of the SDSS counterparts to \FIRST\ sources
also vary systematically with radio flux density (Figs.~\ref{fig-SDSS_mags_colors}a
and \ref{fig-SDSS_mags_colors}b).  Changes in both of these quantities
result from the transition from AGN-powered radio emission
for the brightest radio sources to star-formation-powered radio emission
for milliJansky radio sources.  Around 1~Jy, the supermassive black hole ``monsters''
dominate the counterparts for both stellar objects and galaxies;
objects appear as blue, point-like quasars when nuclear emission also dominates
the optical, while radio galaxies have similar radio fluxes but
are 10 times fainter in the optical.
As the radio flux declines from 1~Jy to 1~mJy,
quasars become both fainter and redder, mainly because the population is
shifting to higher redshift.  The magnitudes and colors of galaxies change
little from 1~Jy to 100~mJy because the radio galaxy counterparts to
bright radio sources are already close to the SDSS detection limit.
However, below 100~mJy the median magnitudes of \FIRST-selected galaxies actually
get brighter as the radio flux decreases.  This somewhat unexpected result is
caused by the \FIRST\ detection of nearby ($z<0.5$) star-forming
galaxies.  These objects are low luminosity radio sources but are bright optical sources.
As star-forming galaxies take over the sample compared with low-luminosity
radio galaxies, the median galaxy brightness increases. The galaxy colors
initially get redder because of the declining contribution of blue nuclear emission,
and then around 4 mJy the galaxy colors begin to become bluer as nearby
galaxies with high star-formation rates are detected.

\bigskip
\bigskip
\section{Conclusions: Lessons for future radio sky surveys} \label{conclusions}

Rather than reiterate our results in a summary, we draw conclusions from the two-decade experience of the \FIRST\ survey that may be of use for the next-generation projects to map the radio sky. We include remarks on scheduling, the continuing usefulness of uniform sky surveys, and the all-important matter of angular resolution in radio source identification, concluding with brief remarks on the value of a JVLA sky survey.

\subsection{Scheduling}

A priority for any sky survey is uniformity. This is best achieved
when the hardware, software, and researchers change little over the
course of the project. Our original proposal to survey the radio
sky with the VLA suggested arranging for a special, hybrid array to
yield both high resolution (and thus, high astrometric accuracy)
and high surface brightness sensitivity (to detect nearby galaxies),
and devoting six months to the project, after which normal VLA
operations would resume. In the event, two separate surveys were
conducted, consuming approximately nine months of observing time;
in the case of the \FIRST\ survey, this was spread over eighteen
years. As noted earlier, the allocated observing windows were not
optimally matched to the sky area to be covered, meaning the
observations often had to be carried out off the meridian
(which would have been optimal). Queue scheduling, the antithesis
of careful planned and optimized observing windows, should
be avoided at all costs for survey observations.

Hardware and software changes accumulate the longer a survey takes.
Examples from the \FIRST\ experience are briefly summarized in
\S\ref{history}; the changes to the researchers over two decades
are best left to the reader's imagination.

\subsection{The legacy value of uniform sky surveys}

The papers of record describing the two VLA surveys --- \cite{becker1995}
and \cite{condon1998} --- have received over 4000 citations;
both papers recorded their highest annual citation rates in 2014,
nineteen and sixteen years after their publication, respectively.
Over 6~million snapshot images have been
downloaded from the \FIRST\ survey web site alone. If 
those queries replaced three-minute snapshot observations, the
observing time saved amounts to 75 times the total time invested in
the survey by the VLA.  And the number of images accessed continues to
increase with time. Over the past two years (2012 June through
2014 June), more than 2.5 million image cutouts were extracted
from our image server by 2500 different users around the world.
Every 20 seconds our server delivers a cutout that is the
the equivalent of a 1-minute snapshot observation with the
current JVLA receivers; in 3 weeks our server distributes snapshots
with a combined exposure time equivalent to the entire 4000 hours
of VLA time that was allocated for the \FIRST\ survey.
The investment of observing time in
the \FIRST\ survey continues to pay dividends to the astronomical
community.

\subsection{Angular resolution and source identification}\label{resolution}

Radio surveys require high angular resolution in order to have positions sufficiently
accurate to obtain source identifications with objects at other wavelengths.
Optical and infrared counterparts of even relatively bright radio
sources are faint: e.g., only 33\% of \FIRST\ radio sources
have an optical counterpart bright enough to
be detected in SDSS.  An important corollary to the fact that radio source identifications are faint is that potential counterparts are dense
on the sky, and accurate radio positions are required to confidently
associate the radio and optical objects.

\subsubsection{Signal-to-noise ratios and source positions}

Can deeper radio observations at low resolution be a
substitute for higher resolution observations?
A common argument of advocates for lower resolution
surveys is that as the
signal-to-noise ratio (SNR) increases,
the positions of catalog sources improve. Consequently
one does not really need high
resolution to do optical identifications.
The prediction of this \textit{SNR model} is that as the
flux density increases, the positional error
will decrease as $1/\SNR$, allowing the optical counterpart to be
confidently matched. Specifically, the NVSS description \citep{condon1998}
gives this formula for
the noise in RA or Dec for point sources:
\begin{equation}
\sigmaoned = \theta / (\SNR \sqrt{2\ln 2}) \quad .
\label{eqn-sigmaoned}
\end{equation}
Here $\theta$ is the resolution FWHM (45\arcsec\ for NVSS) and
$\SNR$ is the signal-to-noise ratio. The median NVSS rms noise for
objects that match \FIRST\ sources is 0.47~mJy. Note that this noise equation
already has been increased by an empirical factor of $\sqrt{2}$ compared
with the theoretical equation ``to adjust the errors into agreement
with the more accurate \FIRST\ positions''
(quoting the NVSS catalog description\footnote{\url{http://www.cv.nrao.edu/nvss/catalog.ps}})
This predicts $\sigmaoned \sim 7.6$~arcsec at the catalog detection
limit ($\SNR=5$) and $\sigmaoned \sim 1$~arcsec at a flux density of 18~mJy.

The positional scatter in Eq.~(\ref{eqn-sigmaoned}) is a 1-dimensional uncertainty, giving the error
in either RA or Dec.  In a 2-dimensional distribution, many values will
scatter outside the 1-sigma circle.  The 
90\% confidence separation limit $\sigmaninety$, which is typically more appropriate
for catalog matching, is a constant factor
$\sqrt{2\ln10}$ times larger than $\sigmaoned$:
\begin{equation}
\sigmaninety = {\theta \over \SNR} \sqrt{\ln 10 \over \ln 2} \quad .
\label{eqn-sigmaninety}
\end{equation}
With this increase it is necessary for the NVSS flux density to exceed
40~mJy ($\SNR=85$) to reduce the predicted separation error to 1~arcsec.

\begin{figure*}
\centering
\includegraphics[width=0.8\linewidth]{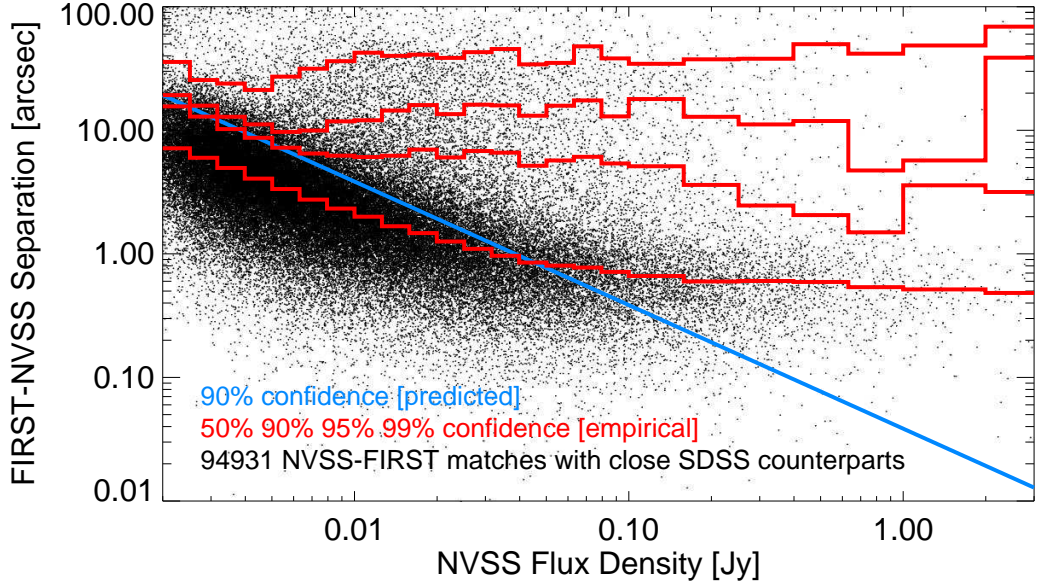} 
\caption{Position difference between NVSS and \FIRST\ positions as a
function of NVSS flux density.  The sample includes only objects
that have a close SDSS counterpart to the \FIRST\ source position
(within 0.7\arcsec).  {\sl Blue line:} Theoretical 90\% confidence
separation limit computed using the SNR as in Eq.~(\ref{eqn-sigmaninety}).
{\sl Red
histograms:} (from bottom to top) the 50, 90, 95, and 99\% confidence limits, computed by determining the actual separations in each bin. 
While the 50\% curve behaves approximately as expected, the tail of the distribution
is clearly non-Gaussian and has many more distant outliers than expected based on
the predicted 90\% curve.
\label{fig-nvsssep}}
\end{figure*}

\begin{figure*}
\includegraphics[width=0.5\linewidth]{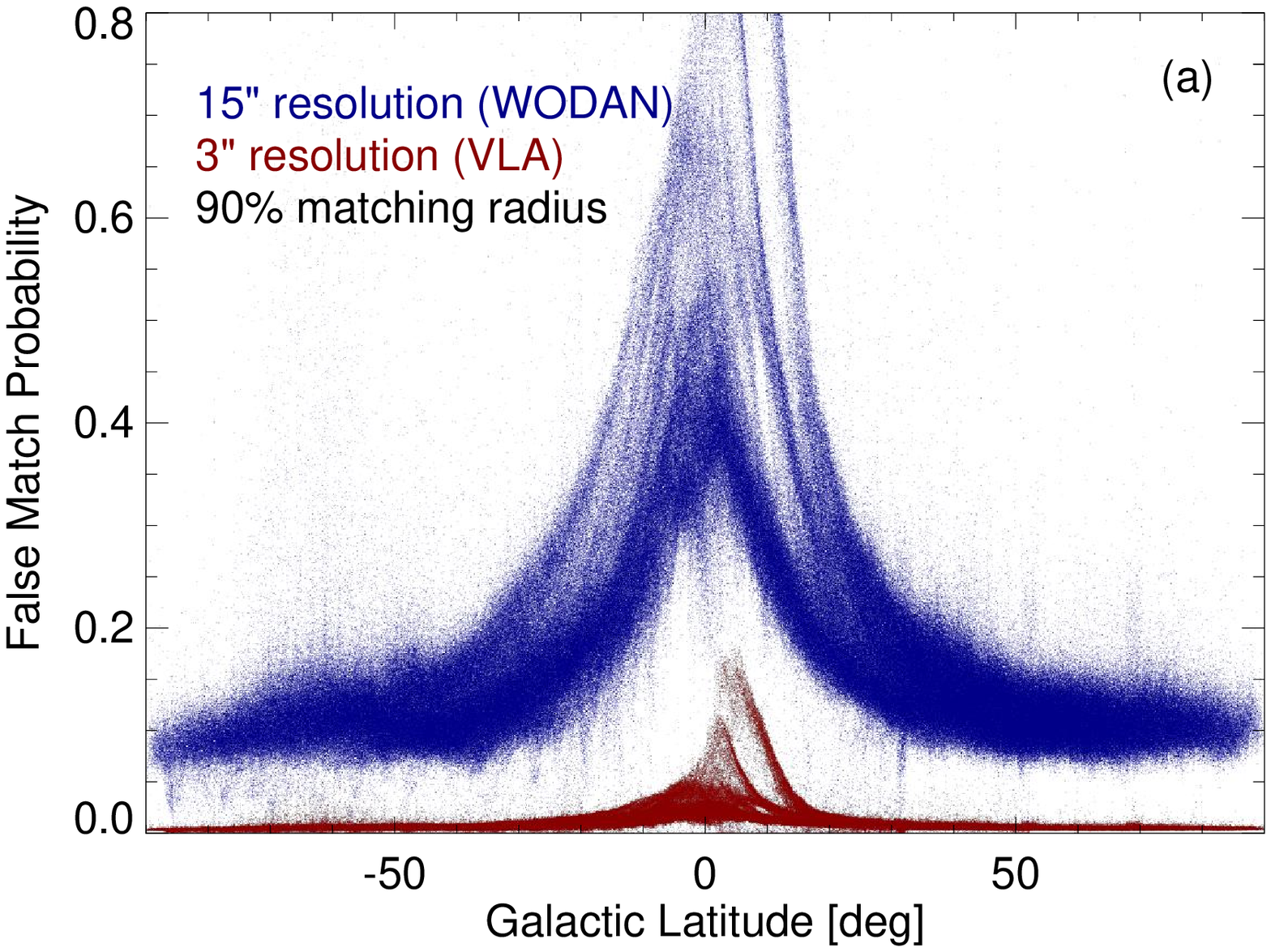}
\includegraphics[width=0.5\linewidth]{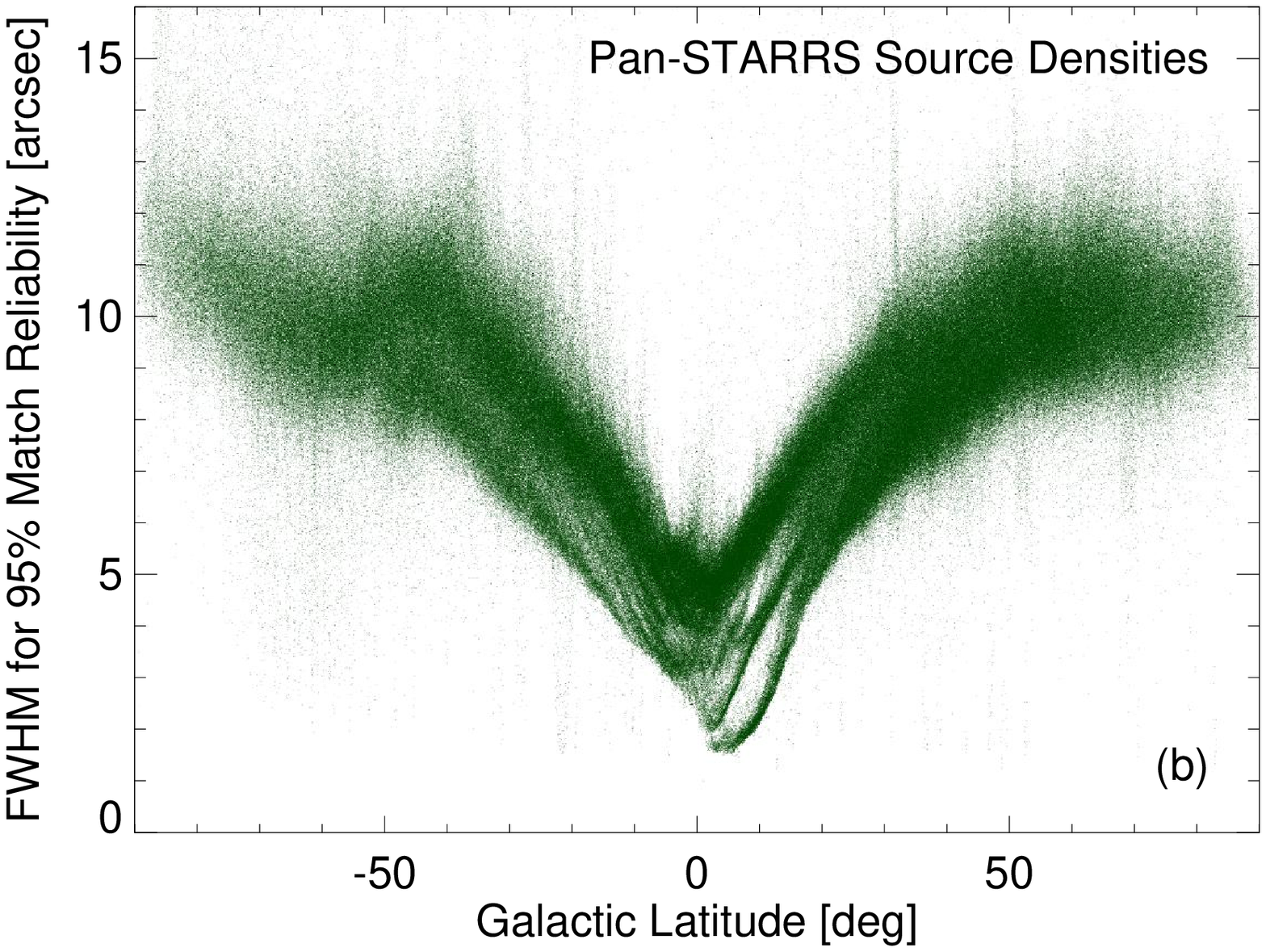}
\caption{{\sl Left:} Probability of a false match in Pan-STARRS as a function of Galactic latitude.
{\sl Right:} FWHM resolution required to achieve 95\% cross-match reliability.
The actual density of PS1 objects was used to calculate the likelihood
of a false counterpart within the 95\% confidence radius.
The bands come from filamentary structure in the dust absorption, which
leads to regions of low and high object density in the Pan-STARRS catalog.
The left panel shows probabilities both for
the WODAN $15\arcsec$ resolution (blue) and the VLA S-band $3\arcsec$ resolution (red).
Over most of the extragalactic sky $\sim10$\% of WODAN-PS1 cross-matches will be
chance coincidences, compared with $<1$\% of the VLA-PS1 matches.  The VLA
positions are sufficient for identifications even close to the Galactic plane.
The right panel shows that in the
extragalactic sky ($|b|>30^\circ$) a resolution better than 11\arcsec\ is required.
WODAN does not have the required resolution; ASKAP-EMU just reaches this
limit but will not cover most of the northern sky.  The
VLASS S-band survey easily meets this requirement.
\label{fig-ps1false}}
\end{figure*}

\subsubsection{Does the SNR model work for NVSS?}

The above positional accuracy applies to perfect
point sources (and perfect data). But how well does it work for
real data?
We can assess the accuracy of Eqn.~(\ref{eqn-sigmaninety})
using a
comparison of the \FIRST\ and NVSS data.

We selected a sample of all the \FIRST\ sources that have an SDSS match
within 0.7 arcsec and that have an NVSS match within 100 arcsec.
We restricted the sample to sources with \FIRST\ peak flux densities
greater than the 2.5~mJy NVSS detection limit.
For all these $\sim95,000$ sources, we computed the distance to the
nearest NVSS source. The important thing about this sample is that
the \FIRST\ source matches the optical source position. That means
that if NVSS is to identify the same counterpart, it needs to
have a position close to the \FIRST\ source position. There may be several \FIRST\
source components associated with a single NVSS source, but only
the \FIRST\ sources that match optical counterparts are included.

How do the positional errors in Eqn.~(\ref{eqn-sigmaninety}) compare
with reality?  Fig.~\ref{fig-nvsssep} shows the position
distance between the NVSS and \FIRST\ positions as a function of
the NVSS flux density.  The positional differences
decrease as expected as the flux densities increase.  The blue line shows the
90\% confidence separation limit $\sigmaninety$ from
Eq.~(\ref{eqn-sigmaninety}), simply assuming that all objects are point
sources with the median NVSS rms value.  Between 1 and 100~mJy, this line generally
tracks the decline in positional differences as the flux density
increases; however, there are still many points above the line.

The red histogram shows the empirical 50, 90, 95, and 99\% confidence separation
limits as a function of flux density, computed by determining the relevant
percentile of the actual separations in each bin. The computed separation
has been corrected for the effects of chance nearby NVSS associations.
\textit{The actual
90\% confidence radius shows no improvement in the positions for
flux densities greater than 4~mJy, and it is
much larger than the predicted 90\% curve.}


This empirical distribution does not look like the theoretical
distribution. Remember that there is an optical counterpart near
zero separation in these plots for every source. To find 90\% of those
counterparts using the NVSS positions, it is necessary to use a matching
radius of approximately 7 arcsec ($\sim0.15\,\theta$) even for sources that are 100
times the rms noise level. The theoretical SNR model predicts that
the positions for such bright sources ought to be much more accurate than that
($\sigmaninety = 0.8\arcsec$).  To include 95\% of the
counterparts requires a matching radius of 15\arcsec\ ($\sim0.3\,\theta$), while
finding 99\% of the counterparts requires matching out to 39\arcsec\ ($\sim\theta$).

\subsubsection{Why are the low-resolution positions so inaccurate?}

Why are the inaccuracies in the positions so much greater than the SNR model
predictions? Real radio sources are not symmetrical objects. They
have lobes, jets, and cores; star-forming galaxies have spiral arms.
And there can be confusion when multiple radio sources get mixed
together in a low-resolution beam. A low-resolution survey does
indeed give a very accurate measurement of the mean flux-weighted
position as the SNR increases. \textit{However, the flux-weighted centroid
is often not where the optical counterpart lies.} In many cases the
counterpart is associated with some sharp structure within the radio
source, and that structure may be far from the flux-weighted center.

\begin{figure}[b] 
\includegraphics[width=\linewidth]{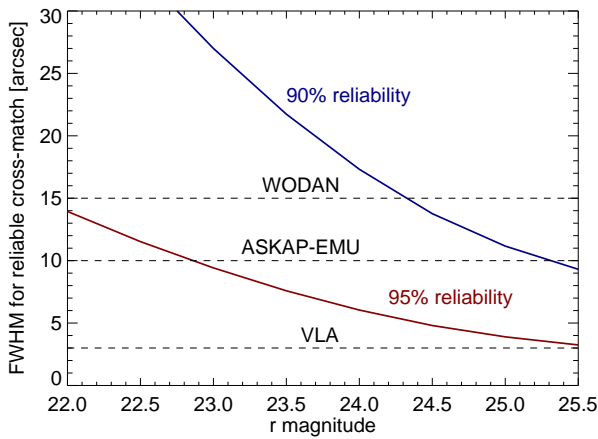}
\caption{
FWHM resolution required to achieve reliable cross-matches at fainter magnitudes
using $r$-band galaxy counts.  The curves define the limits for 90\% and 95\% reliable identifications, and the resolutions of
ASKAP-EMU, WODAN, and a VLA 3\arcsec\ resolution survey are shown.  The depths of various $r$-band surveys are 
also shown by the vertical lines (Pan-STARRS, DESI, LSST, HSC). The SKA pathfinders
are at best marginally sufficient for identifications at SDSS/Pan-STARRS depths, 
while a VLA S-band survey is usable with the much deeper surveys.
\label{fig-rcount}}
\end{figure}

\subsubsection{Effect on optical identifications}

This analysis demonstrates that matching at the 45\arcsec\ resolution of NVSS requires
a matching radius of 15\arcsec = 30\% of the NVSS FWHM resolution
to achieve 95\% completeness.
Our experience with the FIRST survey is similar: to get a reasonably
complete list of optical identifications we had to use a matching
radius of 2\arcsec\ $\sim$ 40\% of the FIRST FWHM resolution.  We
argue this is a universal requirement for radio sources, at least
for sources down to the sub-mJy regime: the matching radius that
is required for realistic radio source morphologies is at least 30\% of the
FWHM resolution for 95\% completeness.

The planned Square Kilometer Array (SKA) precursor surveys have relatively
low resolution.
The resolution for WODAN \citep{rottgering2011} is of order $15\times17$~arcsec, while the
resolution for ASKAP-EMU \citep{norris2011} is 10~arcsec.  For 95\% completeness,
WODAN will therefore require
an optical matching radius of 4.8 arcsec and ASKAP-EMU will
require 3 arcsec.
Such large matching radii are a serious problem for optical matching.
The cross-match between
SDSS and \FIRST\ shows that 19\% of \FIRST\ sources have a false (chance)
SDSS counterpart within 4.8\arcsec. For comparison, 33\% of \FIRST\
sources have a true match within 2\arcsec. So one-third of the optical
counterparts at SDSS depth will be false matches when using a
4.8\arcsec\ matching radius. Of course the false rate can be
reduced somewhat
by doing a careful analysis of the likelihood of association as a
function of separation, but when the starting point is a sample
that is contaminated by 33\% false matches, the final list of
identifications is going to be neither complete nor reliable.

The false matching problem will only get worse for deeper
optical/IR data. For example, Pan-STARRS \citep[PS1 --][]{kaiser2002}
goes more than 1 magnitude deeper than SDSS in the red and also goes
into the Galactic plane where
the source density is much higher, demanding better resolution.
We have used the sky density of objects in a preliminary PS1 catalog
to compute the likelihood
of false identifications in PS1 as a function of Galactic latitude.
The left panel of Fig.~\ref{fig-ps1false} compares WODAN to a
VLA S-band (2--4 GHz) B-configuration survey having
3\arcsec\ resolution. For WODAN, 10\% of sources
even in the extragalactic sky ($|b|>30^\circ$) will have a
spurious counterpart in PS1.  For most purposes that is an unacceptable
level of contamination.  In contrast, a VLA survey has only a $<1$\%
contamination rate in the extragalactic sky, and is usable even
quite close to the Galactic plane.

The right panel of Fig.~\ref{fig-ps1false} turns this around and
asks what FWHM resolution is required to achieve a 95\% reliability
($\sim2\sigma$) in matches to the PS1 catalog.  At $|b|>30^\circ$
a FWHM resolution of 11\arcsec\ is required.  That is significantly
higher resolution than WODAN and just at the resolution reached by
ASKAP-EMU, but one
easily satisfied by VLA surveys.  In fact, a VLA survey with a resolution
of 3\arcsec\ has 95\% confident PS1 matches over 99\% of the current
PS1 catalog area, with only the most crowded areas of the Galactic
plane requiring higher resolution.

The next generation of optical/IR surveys will be considerably
deeper than Pan-STARRS.  Fig.~\ref{fig-rcount} shows the resolution
required as a function of magnitude using the $r$-band galaxy counts
from the CFHTLS-D1 1~$\deg^2$ survey \citep{mccracken2003}.  Since
this does not include stars or redder galaxies, it is more optimistic
(and less realistic) at the PS1 limit, but it shows the resolution
required for deeper identifications.  For 90\% reliable identifications,
a 3\arcsec-resolution VLA survey can be used to $r=27.2$, ASKAP-EMU
to $r=25.3$, and WODAN to $r=24.3$.  For 95\% reliable
identifications, the magnitude limits are 24.9 (VLA), 22.1 (ASKAP-EMU),
and 20.7 (WODAN).  The SKA-precursor surveys are usable at
the depth of SDSS and Pan-STARRS in the extragalactic sky, but fall well short of the required
resolution at fainter magnitudes.  The higher resolution VLA survey,
by contrast, is useful at least to $r=26$.

The inescapable conclusion is that we need high resolution to get
the accurate positions required for optical identifications. Deeper
radio imaging is not a substitute for the requisite angular resolution.

\subsection{A JVLA sky survey}

A new generation of radio sky surveys will soon be conducted by
survey arrays such as LOFAR \citep{devos2009}, ASKAP \citep{johnston2008},
MEERKAT \citep{booth2009} and, ultimately, perhaps,
the SKA \citep{schilizzi2010}. 
But the analysis above shows that
none of the three SKA precursors provide the angular resolution
necessary for unambiguous radio source identification
or to resolve complex
source regions. In addition, they operate at wavelengths of 20~cm
or longer (initially, at least). It is thus worth considering whether
the substantially enhanced JVLA might again be used for a sky survey.

The use of the JVLA for $\sim 7$ months could produce a $3\times10^4$~
deg$^2$ survey at 2--4 GHz with an angular resolution two
times that of \FIRST\ and a sensitivity two times greater.
This would produce over three times as many sources (taking into
account the flux density falloff of most sources with frequency),
one-quarter of which would have accurate spectral indices and all
of which would have full Stokes parameters available. The positional
accuracy for the three million sources would be better than
1\arcsec.

\acknowledgments

The success of the \FIRST\ survey is in large measure due to
the generous support of a number of organizations. In particular,
we acknowledge support from the NRAO, the NSF (grants AST 94-19906,
AST 94-21178, AST-98-02791, AST-98-02732, AST 00-98259, and AST
00-98355), the Institute of Geophysics and Planetary Physics (operated
under the auspices of the US Department of Energy by Lawrence
Livermore National Laboratory under contract No. W-7405-Eng-48),
the STScI, NATO, the National Geographic Society (grant NGS No.
5393-094), Columbia University, and Sun Microsystems.

This publication makes use of data products from the Two Micron All
Sky Survey, which is a joint project of the University of Massachusetts
and the Infrared Processing and Analysis Center/California Institute
of Technology, funded by the National Aeronautics and Space
Administration and the National Science Foundation.

Funding for the SDSS and SDSS-II has been provided by the Alfred
P. Sloan Foundation, the Participating Institutions, the National
Science Foundation, the U.S. Department of Energy, the National
Aeronautics and Space Administration, the Japanese Monbukagakusho,
the Max Planck Society, and the Higher Education Funding Council
for England. The SDSS Web Site is \url{http://www.sdss.org/}.
Funding for SDSS-III has been provided by the Alfred P. Sloan
Foundation, the Participating Institutions, the National Science
Foundation, and the U.S. Department of Energy Office of Science.
The SDSS-III web site is \url{http://www.sdss3.org/}.



\begin{thebibliography}{}

\bibitem[Abazajian~et~al.(2009)]{abazajian2009} Abazajian,~K.~N. et al. 2009, \apjs, 182, 543

\bibitem[Ahn et al.(2014)]{ahn2014} Ahn, C.~P., Alexandroff, 
R., Allende Prieto, C., et al.\ 2014, \apjs, 211, 17 

\bibitem[Assafin et al.(2013)]{assafin2013} Assafin, M., 
Vieira-Martins, R., Andrei, A.~H., Camargo, J.~I.~B., 
\& da Silva Neto, D.~N.\ 2013, \mnras, 430, 2797 

\bibitem[Becker~et~al.(1995)]{becker1995} Becker,~R.~H., White,~R.~L., Helfand,~D.~J., 1995, \apj, 450, 559

\bibitem[Booth et al.(2009)]{booth2009} Booth, R.~S., de Blok,
W.~J.~G., Jonas, J.~L., \& Fanaroff, B.\ 2009, arXiv:0910.2935

\bibitem[Condon(1992)]{condon1992} Condon, J.~J. 1992, \araa, 30, 575

\bibitem[Condon et al.(1998)]{condon1998} Condon,~J.~J., Cotton,~W.~D., Greisen,~E.~W., Yin,~Q.~F., Perley,~R.~A., Taylor,~G.~B., Broderick,~J.~J., 1998, \aj, 115, 1693

\bibitem[Condon et al.(2003)]{condon2003} Condon, J.~J., Cotton,
W.~D., Yin, Q.~F., et al.\ 2003, \aj, 125, 2411

\bibitem[de Vos et al.(2009)]{devos2009} de Vos, M., Gunst,
A.~W., \& Nijboer, R.\ 2009, IEEE Proceedings, 97, 1431

\bibitem[Gal-Yam et al.(2006)]{galyam2006} Gal-Yam, A., Ofek,
E.~O., Poznanski, D., et al.\ 2006, \apj, 639, 331

\bibitem[Hodge et al.(2011)]{hodge2011} Hodge, J.~A., Becker, 
R.~H., White, R.~L., Richards, G.~T., \& Zeimann, G.~R.\ 2011, \aj, 142, 3 

\bibitem[H{\"o}gbom(1974)]{hogbom1974} H{\"o}gbom, J.~A.\ 1974, \aaps, 15, 417 

\bibitem[Ivezi{\'c} et al.(2002)]{ivezic2002} Ivezi{\'c}, {\v Z}.,
Menou, K., Knapp, G.~R., et al.\ 2002, \aj, 124, 2364

\bibitem[Johnston et al.(2008)]{johnston2008} Johnston, S., Taylor,
R., Bailes, M., et al.\ 2008, Experimental Astronomy, 22, 151

\bibitem[Kaiser et al.(2002)]{kaiser2002} Kaiser, N., Aussel, H., 
Burke, B.~E., et al.\ 2002, \procspie, 4836, 154 

\bibitem[Kirkpatrick et al.(1983)]{kirkpatrick1983} Kirkpatrick, S.,
Gelatt, C.~D., \& Vecchi, M.~P.\ 1983, Science, 220, 671

\bibitem[Ma et al.(2013)]{ma2013} Ma, C., Arias, F.~E., 
Bianco, G., et al.\ 2013, VizieR Online Data Catalog, 1323, 0 

\bibitem[McCracken et al.(2003)]{mccracken2003} McCracken, H.~J., Radovich, M., Bertin, E., et al.\ 2003, \aap, 410, 17 

\bibitem[McMahon \& Irwin(1992)]{mcmahon1992} McMahon, R.~G., \& Irwin, M.~J.\ 1992, Digitised Optical Sky Surveys, 174, 417

\bibitem[McMahon~et~al.(2002)]{mcmahon2002} McMahon,~R.~G., White,~R.~L., Helfand,~D.~J., Becker,~R.~H., 2002, \apjs, 143, 1

\bibitem[Murthy, Kasif \& Salzberg(1994)]{murthy1994} Murthy, S.~K., Kasif,
S., \& Salzberg, S.\ 1994, arXiv:cs/9408103

\bibitem[Norris et al.(2011)]{norris2011} Norris, R.~P., Hopkins, 
A.~M., Afonso, J., et al.\ 2011, PASA, 28, 215 

\bibitem[Ofek et al.(2010)]{ofek2010} Ofek, E.~O., Breslauer,
B., Gal-Yam, A., et al.\ 2010, \apj, 711, 517

\bibitem[Orosz \& Frey(2013)]{orosz2013} Orosz, G., \& Frey, S.\ 2013, \aap, 553, A13

\bibitem[Reber(1944)]{reber1944} Reber,~G., 1944, \apj, 100, 279

\bibitem[R{\"o}ttgering et al.(2011)]{rottgering2011} R{\"o}ttgering, 
H., Afonso, J., Barthel, P., et al.\ 2011, Journal of Astrophysics and 
Astronomy, 32, 557 

\bibitem[Schilizzi et al.(2010)]{schilizzi2010} Schilizzi, R.~T.,
Dewdney, P.~E.~F., \& Lazio, T.~J.~W.\ 2010, \procspie, 7733,

\bibitem[Schinnerer et al.(2004)]{schinnerer2004} Schinnerer, E.,
Carilli, C.~L., Scoville, N.~Z., et al.\ 2004, \aj, 128, 1974

\bibitem[Schinnerer et al.(2007)]{schinnerer2007} Schinnerer, E., 
Smol{\v c}i{\'c}, V., Carilli, C.~L., et al.\ 2007, \apjs, 172, 46 

\bibitem[Skrutskie et al.(2006)]{skrutskie2006} Skrutskie,~M.~F. et al., 2006, \aj, 131, 1163

\bibitem[Thyagarajan et al.(2011)]{thyagarajan2011} Thyagarajan, N.,
Helfand, D.~J., White, R.~L., \& Becker, R.~H.\ 2011, \apj, 742, 49

\bibitem[White~et~al.(1997)]{white1997} White,~R.~L., Becker,~R.~H., Helfand,~D.~J., Gregg,~M.~D., 1997, \apj, 475, 479

\bibitem[White et al.(2000)]{white2000} White, R.~L., Becker,
R.~H., Gregg, M.~D., et al.\ 2000, \apjs, 126, 133

\bibitem[White et al.(2003)]{white2003} White, R.~L., Helfand,
D.~J., Becker, R.~H., et al.\ 2003, \aj, 126, 706

\bibitem[White et al.(2005)]{white2005} White, R.~L., Becker,
R.~H., \& Helfand, D.~J.\ 2005, \aj, 130, 586

\bibitem[White~et~al.(2007)]{white2007} White,~R.~L., Helfand,~D.~J., Becker,~R.~H., Glikman,~E., de~Vries,~W., 2007, \apj, 654, 99

\bibitem[White(2008)]{white2008} White, R.~L.\ 2008, American
Institute of Physics Conference Series, 1082, 37

\bibitem[York~et~al.(2000)]{york2000} York,~D.~G. at al. 2000, \aj, 120, 1579

\end{thebibliography}
\end{document}